\documentclass[pra, twocolumn, showkeys, floatfix, nofootinbib]{revtex4}
\usepackage{graphicx}
\usepackage{color}
\usepackage{amsmath, amsfonts, amssymb, bm}
\usepackage{tikz}
\usepackage{slashed}
\newcommand{\dd}{\text{d}}
\begin{document}
	\title{Bound-free electron-positron pair production in combined Coulomb and constant
	crossed electromagnetic fields: a Schwinger-like process with intrinsic assistance}
	\author{S.~Remme}
	\author{A.~Eckey}
	\author{S.~Villalba-Ch\'avez}
	\author{A.~B.~Voitkiv}
	\author{C. M\"uller}
	\affiliation{Institut f\"ur Theoretische Physik I, Heinrich-Heine-Universit\"at D\"usseldorf, Universit\"atsstra{\ss}e~1, 40225 D\"usseldorf, Germany}
	\date{\today}
	\begin{abstract}
	The bound-free channel of electron-positron pair production by a highly charged bare ion in the presence of a strong constant crossed electromagnetic field is studied. To calculate the pair production rate, two different methods are applied and compared with each other: (i) a quasiclassical tunneling theory and (ii) a strong-field approximation, both equipped with appropriate Coulomb correction factors. The resulting rate, which depends nonperturbatively on both the Coulomb field of the ion and the constant crossed field, is calculated in a broad range of applied field strengths and nuclear charge numbers. Its functional form resembles the rate for a dynamically assisted Schwinger-like process, with the assistance being provided by the atomic binding energy of the created electron.
	\end{abstract}

	\maketitle
	
	\section{Introduction}
	It constitutes a fundamental property of quantum electrodynamics that certain classes of strong electromagnetic fields can extract electron-positron pairs from the quantum vacuum. In the case of static or slowly varying fields, an electric field strength of the order of the critical value $F_c=m^2c^3/(e\hbar)\approx 1.3\times 10^{16}$\,V/cm is required to reach a sizeable pair yield \cite{Sauter}. Here, $m$ is the electron mass, $e$ the elementary charge, $c$ the speed of light, and $\hbar$ the reduced Planck constant. Notably, pair production in a constant electric field $F_0$ follows the famous Schwinger rate $\mathcal{R} \sim \exp(-\pi F_c/F_0)$, exhibiting a manifestly nonperturbative field dependence \cite{Schwinger}. Similar dependencies are obtained for pair production processes in other electromagnetic field configurations \cite{Ritus-Review, Review1, Review2, Review3, Review4}. 
	
	Due to the ongoing technological progress in high-intensity laser facilities \cite{ELI, CoReLS, FACET, Gemini, LUXE, CALA}, pair production in strong laser fields has been of special interest in recent years. Since a plane electromagnetic wave cannot produce pairs by itself, combinations with other fields -- such as a Coulomb field or a $\gamma$-ray photon field -- have been considered, starting with seminal studies in the 1960s \cite{Reiss, Nikishov-Ritus, Yakovlev, Ritus-trident, Narozhny-Nikishov}. The recent interest in strong-field pair production has been stimulated by the first observation of laser-induced electron-positron pair production that was accomplished at the Stanford Linear Accelerator Center, where a highly relativistic electron beam was brought into collision with an intense laser pulse \cite{SLAC}. The experiment operated in a few-photon regime at a moderate value $\eta\lesssim 1$ of the dimensionless intensity parameter $\eta = ea/(mc^2)$, where $a$ denotes the amplitude of the laser vector potential, and observed a production rate of power-law form $\mathcal{R} \sim \eta^{10}$.
	
	A similar setup is conceivable for pair production in combined nuclear Coulomb and intense laser fields (so-called nonlinear Bethe-Heitler process), when a highly relativistic bare ion -- with Lorentz factor $\gamma\gg 1$ -- collides with a counterpropagating laser wave. In such collisions, the relevant laser field strength is strongly enhanced by a Lorentz boost factor $\approx 2\gamma$ to the rest frame of the ion, which facilitates to approach the critical field scale. The free-free channel of the nonlinear Bethe-Heitler process, where both the electron and the positron are created in continuum states, has been studied thoroughly in different $\eta$-regimes and for various laser polarizations \cite{Yakovlev, Mittleman, Avetissian, MVG-PRA2003, Sieczka, Krajewska-PRA2006, Milstein, Kuchiev, DiPiazza-PLB}. Noteworthy, when $\eta\gg 1$ and the (Lorentz-boosted) laser field $F_0$ remains far below $F_c$, the production rate $\mathcal{R} \sim\exp(-2\sqrt{3}F_c/F_0)$ attains a Schwinger-like form \cite{Milstein}. In this situation, the pair formation length $\ell\sim mc^2/(eF_0) \sim \lambda_{\text{L}}/\eta$ is much smaller than the wavelength $\lambda_{\text{L}}$ of the laser field, so that the latter appears quasistatic.\footnote{This argument forms the basis of the locally constant field approximation which is often used in numerical calculations of strong-field QED processes in slowly varying fields of large $\eta$ value; see, e.g., Sec.5.1 in \cite{Review4} and \cite{LCFA1, LCFA2, LCFA3}. A complementary approach is the locally monochromatic approximation which applies to many-cycle laser fields whose pulse envelope changes slowly; see Sec.5.2 in \cite{Review4} and \cite{LMA1,LMA2}.} Nonlinear Bethe-Heitler pair creation was also analyzed in pulsed \cite{Lebed, Krajewska-PRA2013, Krajewska-NJP} and two-color \cite{Krajewska-PRA2012, Augustin-PRA} laser fields. Moreover, spin effects were studied \cite{Tim-Oliver} as well as pair creation by two atomic nuclei in a laser field \cite{diatomic}, or by a nucleus in a standing laser wave \cite{Grobe-phase}.
	
	When pairs come into existence in the presence of a nucleus, also bound-free pair production can occur, where the electron is created in a bound atomic state, while the positron is emitted into the continuum. In the Dirac-sea picture, the process can be seen as the excitation of an electron from the negative-energy continuum to a discrete positive-energy state bound to the atom \cite{Agger, Sommerfeldt}. This channel of the nonlinear Bethe-Heitler process has so far been studied solely in a domain of small intensity parameters $\eta\ll 1$, where the pair production proceeds in a perturbative few-photon regime \cite{MVG-PRL, MVG-PRA2004, Deneke}. Relativistic ions with $\gamma\approx 50$ colliding with intense x-ray laser fields of about 10\,keV photon energy were assumed, leading to MeV photon energies and, thus, very rapid laser oscillations in the nuclear rest frame. Accordingly, power-law rate dependencies on the intensity parameter were obtained for both circularly \cite{MVG-PRL, MVG-PRA2004} and linearly \cite{Deneke} polarized laser fields. Bound-free pair production in this regime has also been studied by a nucleus located in a standing laser wave, leading to pronounced phase effects \cite{Grobe-phase}.
	
	The present paper is devoted to the complementary regime of field-induced bound-free pair production, where the process occurs in a (quasi)static setting, corresponding to a slowly varying laser field. In this scenario, an exponential Schwinger-like behavior of the pair production rate is expected to arise, as will first be argued qualitatively within the theoretical framework that was used in the earlier studies \cite{MVG-PRL, MVG-PRA2004, Deneke}. In the main part of our considerations, we shall apply two different methods to calculate the bound-free pair production rate, which rely on either (i) a quasistatic tunneling theory or (ii) a strong-field approximation, and account for Coulomb corrections in an appropriate way. Our approaches take advantage of the close connection between strong-field pair production and relativistic ionization \cite{Brabec, BrabecLett, Alex, Klaiber_momentumshift1, ImaginaryTime1,ImaginaryTime2,ImaginaryTime_QM, TunnelMPI, Nikishov1, Nikishov2, Reiss_JOpt}. The external field will be modeled by a constant crossed electromagnetic field, wherein a highly charged bare ion is located. Total pair production rates in a wide range of external field strengths and nuclear charge numbers are presented. In particular, we will show how the bound-state nature of the created electron affects the Schwinger-like rate dependence and reveal a mechanism of ``intrinsic'' assistance, where the binding potential of the bound-state electron plays a similar role as the photon energy of a superposed high-frequency field in the process of dynamically assisted (free-free) pair production \cite{Schuetzhold, Schwinger1, Schwinger2, Schwinger3, Schwinger4, Schwinger5, DiPiazza-PRL, Augustin-PLB}. 
	
	We note that the obtained rate expressions in a constant crossed field can be used in future studies to describe bound-free pair production in slowly oscillating laser fields by applying the locally constant field approximation \cite{LCFA1,LCFA2,LCFA3}. Moreover, in the particular case of a slowly oscillating laser field of circular polarization and constant amplitude, we may expect that the associated rate for bound-free pair production in a Coulomb field will coincide with our result. Indeed, in the case of free-free pair production, the corresponding rates are known to exactly agree with each other \cite{Milstein, Narozhny-Nikishov}.
	
	Our paper is organized as follows. Sec.~\ref{sec:Preliminary_consideration} provides a first estimate for the characteristic exponential dependence of strong-field bound-free pair production. Afterwards, we present our theoretical approaches for bound-free pair production in a constant crossed field, starting with the calculation of the pair production rate using the tunneling theory in Sec.~\ref{subsec:WKB_rate} and the strong-field approximation in the G\"oppert-Mayer gauge in Sec.~\ref{subsec:SFA_rate}. Both derivations are followed by the inclusion of an appropriate Coulomb-correction factor which needs to be multiplied with the bound-free pair production rate in Sec.~\ref{subsec:WKB_Q} and Sec.~\ref{subsec:SFA_Q}. Finally, we illustrate and compare our results for the total pair production rate in  Sec.~\ref{subsec:results_rate} with an emphasis on the exponential dependence and discuss the momentum distribution of the created positron in Sec.~\ref{subsec:results_p}. A conclusion is given in Sec.~\ref{sec:conclusion}.
	
	Atomic units (a.u.) with the reduced Planck constant $\hbar=1$, the electron mass $m=1$, and the elementary charge unit $e=1$ are used throughout. The Minkowski product of four-vectors reads $a\cdot b = a_0b_0 - \boldsymbol{a}\cdot\boldsymbol{b}$, and Feynman slash notation is applied for products with $\gamma$-matrices.
	
	\section{Preliminary consideration}
	\label{sec:Preliminary_consideration}
	Bound-free pair production in combined laser and nuclear Coulomb fields has been treated in a few-photon regime within the strong-field approximation (SFA) \cite{MVG-PRL, MVG-PRA2004, Deneke}. In this section, we consider the process in a circularly polarized laser wave with four-potential of the form 
	\begin{equation}
	\label{A_mu_RG}
	A^\mu (x) = a_1^\mu \cos(k\cdot x) + a_2^\mu \sin(k\cdot x)\,,
	\end{equation}
	with the amplitude four-vectors $a_1^\mu = (0,-a,0,0)$, $a_2^\mu = (0,0,-a,0)$, the wave four-vector $k^\mu = \omega(1,0,0,1)$, and the phase $\varphi_k = k\cdot x$. Within the SFA, the amplitude for bound-free pair production can be formulated as
	\begin{equation}
	\label{S_general}
	\mathcal{S}_{\rm SFA} = \frac{i}{c} \int d^4x\,\overline{\Phi}_{1s}(x) \slashed{V}(x) \Psi_{p,s}(x)\,,
	\end{equation}
	where the electron is assumed to be created in the $1s$ ground state. $V^\mu$ is the four-potential of the laser wave and $\Psi_{p,s}$ denotes the Volkov state of the created positron with asymptotic momentum $p^\mu$ in the continuum, which takes the interaction with the laser wave exactly into account, but disregards the interaction with the nuclear Coulomb field. 
	For the considered circularly polarized laser wave $V^\mu$ is given by $A^\mu$ in Eq.~\eqref{A_mu_RG} when the radiation gauge (RG) is applied. Moreover,
	\begin{equation}
		\label{Volkov-state-RG}
		\begin{split}
			\Psi_{p,s}^{(\rm RG)} = & \sqrt{\frac{c}{p_0}} \left(1 + \frac{\slashed{k} \slashed{A}}{2c(k \cdot p)} \right) v_{s^+} \exp \left(iS^{(+)}\right)
		\end{split}
	\end{equation}
	denotes the corresponding Volkov state of the positron. It contains the classical action 
	\begin{equation}
		\label{action}
				S^{(+)} = p \cdot x + \frac{a}{c (k \cdot p)} \left( p_x \sin\varphi_k - p_y \cos\varphi_k \right) + \frac{a^2\varphi_k}{2c^2k\cdot p}\,.
	\end{equation}
	Finally, $\Phi_{1s} (x) = g(r) \chi_s \exp \left( -i E_{1s} t \right)$ denotes the bound state in the hydrogen-like ion, whose energy amounts to $E_{1s} = c^2 - I_p$ with the atomic binding potential $I_p$ (see Sec.~\ref{subsec:WKB_Q} for details). This state contains the nuclear Coulomb field to all orders, but lacks the influence of the laser field.
 
	By performing a Fourier series expansion of the periodic parts, the amplitude in Eq.~\eqref{S_general} adopts the form
	\begin{equation}
	\label{S_RG_2}
	\mathcal{S}_{\rm SFA}^{\rm (RG)} = \sum_{n\ge n_{\rm min}}^\infty \mathcal{T}_n\, J_n(\rho)\, \delta(q_0 + E_{1s} - n\omega)
	\end{equation}
	with the Bessel function argument $\rho = \frac{\eta c^2}{\omega}\,\frac{\sqrt{p_x^2+p_y^2}}{p_0-p_z}$, $\eta = a/c^2$, and the minimal number $n_{\rm min}$ of laser photons needed to exceed the energy threshold. The $\delta$-function describes the energy conservation in the process, with the laser-dressed energy $q_0 = p_0 + \frac{a^2}{2(p_0-p_z)}$ of the positron.  

	The amplitude \eqref{S_RG_2} consists of an infinite sum of terms. In the quasistatic strong-field parameter regime of interest here ($\eta\gg 1$, $\omega\ll c^2$), a large number of terms with $n\gg 1$ will give relevant contributions. Due to the mathematical properties of the Bessel functions $J_n(\rho)$, the most significant contributions -- which determine the physical characteristics of the process -- stem from those terms where the difference $n-\rho$ is minimal \cite{Hatsagortsyan}. This occurs for the optimum values
	\begin{equation}
	n_0 = \frac{1}{\omega}\left[ (1+\eta^2)c^2 + E_{1s} \right]\,,\ \rho_0 = \frac{\eta^2c^2}{\omega}\,.
	\end{equation}
	Accordingly, $\rho_0 = n_0 - \nu$, where $\nu = \frac{2c^2-I_p}{\omega}\ll n_0$. The Bessel function therefore takes the asymptotic form $J_{n_0}(\rho_0)\sim \exp\Big(-\frac{2\sqrt{2}}{3}\,\frac{\nu^{3/2}}{\rho_0^{1/2}}\Big)$ \cite{Abramowitz}. With $\frac{\nu^{3/2}}{\rho_0^{1/2}} = \frac{F_c}{F}\delta_{\text{bf}}^{3/2}$ and $\delta_{\text{bf}}=2-\frac{I_p}{c^2}$, the exponential dependence of the bound-free pair production rate is obtained as
	\begin{equation}
	\label{R_RG}
	\mathcal{R} \sim \big| J_{n_0}(\rho_0) |^2 
	\sim \exp\left( -\frac{4\sqrt{2}}{3}\,\frac{F_c}{F}\,\delta_{\text{bf}}^{3/2} \right) ,
	\end{equation} 
	in close resemblance to the Schwinger rate. Here, $F=a\omega / c$ is the laser field strength. We emphasize that, throughout our considerations, the amplitude $F$ of the external field is assumed to be substantially smaller than the Coulomb field $F_a$ experienced by the bound $1s$ electron. This condition guarantees that the influence of the nuclear Coulomb field on the bound electron state is much more important than the applied external field, so that this state keeps its main physical properties as a bound state.

	This rate expression \eqref{R_RG} offers an intuitive physical interpretation. The energy gap for pair production, when both particles are created in free states, amounts to $2c^2$, corresponding to a gap parameter $\delta_{\rm bf} = 2$. In case of bound-free pair production, the energy gap is reduced to $\delta_{\rm bf} < 2$ by the binding potential $I_p$, which lowers the exponential suppression of the process and enhances the rate. The effect of the binding potential is very similar to the influence of a weak high-frequency field component that is superposed onto a strong low-frequency field. Free pair production in the latter field configuration has been termed dynamically assisted Schwinger effect \cite{Schuetzhold, Schwinger1, Schwinger2, DiPiazza-PRL, Schwinger3, Schwinger4, Augustin-PLB, Schwinger5}. In the case of strong-field bound-free pair production the assistance is not dynamical, though, but rather provided 'intrinsically' by the reduced eigen energy of the created (stationary) electron state. As a consequence, the process rate depends nonperturbatively on both the laser field and the nuclear Coulomb field.

	It is interesting to note that the same exponential dependence can be obtained through a tunneling consideration where an integration over an approximated momentum of the electron underneath the barrier is performed, such that \cite{DiPiazza-PRL} 
	\begin{equation*}
		\mathcal{R} \sim \exp \left(-2 \int_{0}^{l} p(x) \dd x \right).
	\end{equation*}
	As the reduction of the energy barrier results intrinsically, we approximately set $F l = 2 c^2 -I_p$ and $p(x) = \sqrt{2  \left(2c^2-I_p- Fx\right)}$, where $l$ denotes the tunnel length. Additionally we assume that the momentum vanishes at the tunnel entrance $x=l$ and obtain the exponential in \eqref{R_RG}.

	We emphasize that the result in Eq.~\eqref{R_RG}, while being physically appealing, represents an estimate for the exponential dependence of the bound-free pair production rate only. When the same consideration is carried out for free-free pair production, where $ \rho_0 = \frac{2\eta^2c^2}{\omega}$ and $\nu = \frac{2c^2}{\omega}$, a rate dependence of the form $\sim \exp\big(-\frac{8\sqrt{2}}{3}\frac{F_c}{F}\big)$ follows \cite{Hatsagortsyan}, which differs from the exact exponential dependence given by $\sim \exp\big(-2\sqrt{3}\,\frac{F_c}{F}\big)$ (see, e.g., Eq.~(41) in \cite{Milstein}).

	In principle, the rate for strong-field bound-free pair production within the present approach could be evaluated more accurately by numerical means. It is, however, well known from studies of strong-field ionization that SFA calculations within the radiation gauge (respectively the velocity gauge in the nonrelativistic domain) do not lead to quantitatively convincing results \cite{Bauer}. Instead, two alternative approaches have proven successful for relativistic strong-field ionization: (i) a tunneling theory based on the WKB approximation \cite{BrabecLett, Brabec}, and (ii) the SFA within G\"oppert-Mayer gauge (which is the relativistic counterpart to the length gauge) \cite{Klaiber_momentumshift1, Alex}. In the following Sec.~III and Sec.~IV we will apply both these approaches -- with appropriate adjustments -- to the process of bound-free pair production.
	
	\section{Bound-free pair production in WKB tunneling theory}
	\label{sec:WKB}
	
	In the following section, we outline the bound-free pair production in a combined laser and Coulomb field using the tunneling theory based on a WKB tunneling approximation. For the laser field we consider a constant crossed field (CCF) which represents the infinite-wavelength limit of a plane wave. This field model is often applied in the $\eta \gg 1$ regime, where the pair formation length $\sim \lambda_{\text{L}} / \eta$ is much smaller than the scale of field variations which is given by the laser wave length $\lambda_{\text{L}}$. The electric field  $\mathbf{F}$ is oriented along the $x$ axis and the magnetic field $\mathbf{B}$ along the $y$ axis, with equal amplitudes $|\mathbf{F}|= |\mathbf{B}| = F$.
	
	\subsection{Pair production rate in WKB-theory}
	\label{subsec:WKB_rate}
	
	For the derivation of the bound-free pair production rate with a WKB-approach, we use the calculation of the relativistic ionization rate in \cite{Brabec, BrabecLett} as orientation and adjust it to the pair production process.
	
	In case of bound-free pair production, the electron tunnels from the negative energy continuum into the bound $1s$ state. Its equation of motion consequently reads
	\begin{equation}
		\label{eq:equation of motion}
		\frac{\dd \mathbf{p}}{\dd t} = -F \mathbf{e}_x - \frac{F}{c} \mathbf{v} \times \mathbf{e}_y ,
	\end{equation} 
	where  $\mathbf{v}$ is the electron velocity and $c = 137 \ \text{a.u.}$ is the speed of light. Note that even though  we consider the pair production process in a combined laser and Coulomb field, the Coulomb field is disregarded in the electron's trajectory. Its influence will be incorporated at a later stage (see Sec.~\ref{subsec:WKB_Q}).

	As the electron initially starts in the negative energy continuum, the energy is defined by $E= - \gamma c^2$, where $\gamma = \frac{1}{\sqrt{1- (v/c)^2}}$ denotes the Lorentz factor. During the tunneling process it moves from the negative energy continuum into a region with positive energy. As we will show in the following, we obtain an expression for the Lorentz factor which changes its sign during the process such that a separate consideration of both regions is not necessary. Thus, the sign change of the energy is obtained automatically within this calculation. Similarly, we also define the relativistic momentum with an additional minus sign in front of the Lorentz factor $\mathbf{p}= - \gamma \mathbf{v}$ which ensures that velocity and kinetic momentum are directed in the same direction for positive energy states (where $-\gamma$ undergoes a sign change and becomes positive).
	
\begin{figure}[t]
		\centering
		\begin{tikzpicture}[scale =0.8]
			\centering
			\node at (9,6) {$\mathbf{F} = F \mathbf{e}_x$};
			\draw[stealth-,black,  thick] (11,5.5) -- (7,5.5) ;
			\draw[black, thick] (3,1) -- (7,5) node[pos=-0.05] {$+mc^2$};
			\draw[black, thick] (7,1) -- (11,5) node[pos=-0.05] {$-mc^2$};
			\draw[black, thick] (5.35,3) -- (5.85,3);
			\draw[black, thick, text width=1cm, align=center] (9,2.75) -- (9,3.25);
			\node[black, thick, text width=1cm, align=center] at (9.3,2) {$t=0$ $x=l$ $u=0$};
			\draw[-stealth,black, thick] (10,3) -- (6,3);
			\filldraw[black, text width=2cm, align=center] (5.6,3) circle (3pt) ;
			\node[black, thick, text width=1.1cm, align=left] at (5.6,2) {$t=t_0$ $x=0$ $u=u_0$};
		\end{tikzpicture}
		\caption[justification=justified]{Schematic illustration of the pair production tunneling process. Initially the electron is in the negative energy continuum. $t=0, \ x=l \ \text{and} \ u=0$ denotes the tunnel entrance, as indicated, and $t=t_0, \ x=0 \ \text{and} \ u=u_0$ is associated with the final bound state.}
		\label{fig:schematische-darstellung-gfpp}
	\end{figure}
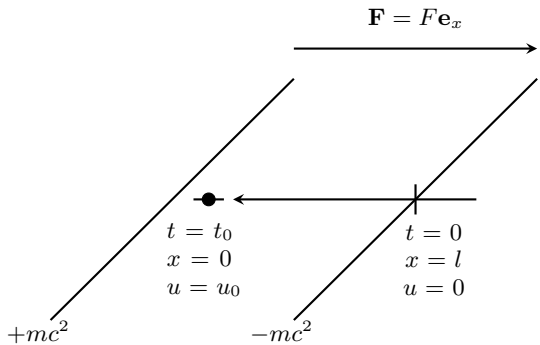

	For illustration, Fig.~\ref{fig:schematische-darstellung-gfpp} shows the electron's tunneling process from the negative energy continuum into the bound $1s$ state. 
	The electron's trajectory is obtained from the equation of motion \eqref{eq:equation of motion}, analogously to the calculation of the ionization trajectory in \cite{LandauF, ImaginaryTime1, Brabec, BrabecLett}. We consider a parametric form of the trajectory, depending on a new variable $u$, where  $u=0$ denotes the tunnel entrance and $u=u_0$ corresponds to the electron's bound state (compare Fig.~\ref{fig:schematische-darstellung-gfpp}).
	
	Compared to the relativistic ionization, the initial conditions need to be adapted to the negative energy in the initial state, such that $\text{Im} \left(\mathbf{r} \left(0\right)\right) = 0$, $\mathbf{r} \left(u_0\right) = 0$,  $\text{Im} \left(\dot{\mathbf{r}} \left(0\right)\right) = 0$ and $\varepsilon = - \frac{1}{\sqrt{1-\dot{\mathbf{r}} \left(u_0\right)^2 / c^2 }}$. Here, $\varepsilon$ is the scaled electron's energy $\varepsilon = \frac{E_{1s}}{c^2} = \sqrt{1- \left(\frac{Z}{c} \right)^2}$ with the bound state energy $E_{1s}$ and the nuclear charge $Z$. The resulting parametric form of the trajectory reads
	\begin{equation}
		\label{trajectory}
		\begin{split}
		x(u) & = \frac{c^2}{2F \lambda} \left[ u_0^2 - u^2 \right] , \\
		y(u) & = 0 , \\
		z(u) & = \frac{ic^2}{6F \lambda} \left[ u_0^2 - u^2 \right]u , \\
		t(u) & = \frac{ic}{2F\lambda} \left[\left( \lambda^2+1 \right) u- \frac{1}{3} u^3 \right] .
		\end{split}
	\end{equation}
	For the sub barrier trajectory of bound-free pair production, we consider $0 \leq u \leq u_0 = \sqrt{3 \left(  \lambda^2 -1 \right)} $, where $\lambda$ results from the initial condition $\varepsilon = - \frac{1}{\sqrt{1-\dot{\mathbf{r}} \left(u_0\right)^2 / c^2 }} = - \frac{1 + \lambda^2 - u_0^2}{2 \lambda}$ as 
	\begin{equation}
		\label{lambda_pp}
		\lambda = \frac{1}{2} \left(\sqrt{\varepsilon^2 + 8} + \varepsilon \right).
	\end{equation}
	Note, that for relativistic ionization, $\lambda_I$ results as $\lambda_I = \frac{1}{2} \left(  \sqrt{\varepsilon^2 + 8} - \varepsilon \right)$ \cite{Brabec, BrabecLett} and only differs from the $\lambda$ of pair production by one minus sign. However, this sign leads to considerably larger values of $\lambda$ which further results in a larger tunneling width and consequently smaller pair production than ionization rates.
	
	Following from the parametric form of the trajectory, the total classical energy and momenta are found to be $E(u) = - \gamma (u) c^2= - \frac{1+ \lambda^2 -u^2}{2 \lambda} c^2$, $p_{xc} (u) = - i \frac{cu}{\lambda} = - ic \sqrt{1+(p_{zc}/c)^2- \left(- \varepsilon + Fx / c^2 \right)^2}$ and $p_{zc} (u) = p_{z0} -\frac{1}{c} Fx$, where $p_{z0} = \frac{c}{4} \left( 3 \varepsilon + \sqrt{\varepsilon^2+8} \right)$ denotes the electron's canonical momentum in $z$-direction. Note that the momentum is imaginary as the motion underneath the barrier is classically forbidden. To further include the quantum mechanical uncertainty of trajectory and momenta, a variation of the momenta is introduced such that $p_x = p_{xc}, \ p_y = \delta p_y \ \text{and} \ p_z = p_{zc} + \delta p_z$. 
	
	For the derivation of the pair production rate, we start by assuming that the electron in the final state is bound in a short-range $\delta$-potential. The Coulomb potential will later be included by the Coulomb correction factor in Sec.~\ref{subsec:WKB_Q}.  The bound $1s$ state of the short-range Dirac atom is given by \cite{Brabec}
	\begin{equation}
		\label{eq: wavefunction bound state short range}
		\begin{split}
		\Psi_{\text{sr}}(\mathbf{r}) = & \frac{N \sqrt{\kappa}}{r} \ \exp \left(- \kappa r\right) \\
		& \times \left(1  , 0  , \frac{\kappa + 1/r}{1+ \varepsilon} \cos \vartheta  , \frac{\kappa + 1/r}{1+ \varepsilon} \sin \vartheta  \text{e}^{i \varphi } \right)
		\end{split}
	\end{equation}
	where $\kappa = c \sqrt{1 - \varepsilon^2}$, and $\vartheta$ and $\varphi$ are the spherical angles. $r$ is the radial distance but will be considered in cartesian coordinates in the following. The normalization constant $N$ of the short-range wave function is left unspecified as for a $\delta$-potential the wave function cannot be normalized due to divergences \cite{Brabec}. $N$ will cancel out when including the Coulomb correction. For the bound state it is assumed that the effect of the crossed field is negligible.
	
	Additionally, the electron underneath the potential barrier can be described by a quasi-classical WKB wave function of the form \cite{LandauQM}
	\begin{equation}
		\label{eq:quasi-classical wave function}
		\Psi_{\rm qc} (p,x) = \frac{C}{\sqrt{|p|}} \exp \left( i  \int_{x}^{x_1} p(x') \dd x'  + i \frac{\pi}{4}  \right).
	\end{equation}
	It is assumed that a matching position $x= x_1$ with $a_0 \ll x_1 \ll l$ underneath the barrier exists where both the bound state \eqref{eq: wavefunction bound state short range} and the quasi-classical wave function describe the electron's state.  Here, $a_0 = 1/Z$ is the Bohr-radius of a hydrogen-like ion with nuclear charge $Z$ while $l= r(u=0) = x(u=0) = \frac{u_0^2}{2F \lambda} c^2$ denotes the tunnel length. Eq.~\eqref{eq:quasi-classical wave function} can consequently be set equal to the Fourier transformation in $y$ and $z$ of Eq.~\eqref{eq: wavefunction bound state short range}. As the Fourier transformation is needed at the matching point we exploit that here $x \gg y, z$; $\kappa \gg p_y, p_z$ and $x \gg \frac{1}{\kappa}$. This leads to \cite{Brabec}
	\begin{equation}
		\label{eq:Fouriertransform qc}
			\tilde{\Psi}(\tilde{p}_x ,x) = \frac{2 \pi \sqrt{\kappa}}{\tilde{p}_x} \text{e}^{-x \tilde{p}_x}
	\end{equation}
	with $\tilde{p}_x = \sqrt{\kappa^2 + p_y^2 + p_z^2}$. Setting Eq.~\eqref{eq: wavefunction bound state short range} and Eq.~\eqref{eq:Fouriertransform qc} equal at $x=x_1$ under the assumption that $\tilde{p}_x \approx \tilde{p}_x (x=x_1) \approx \sqrt{\kappa^2 + p_{z0}^2}$ and $p \approx - i \tilde{p}_x$, the wave function results in \cite{Brabec}
	\begin{equation}
		\label{eq: wave function}
		\Psi_{\text{qc}} (p, x) = \frac{2 \pi N K }{\sqrt{|p|}} \exp \left(- i \int_{0}^{x} p \ \dd x \right)
	\end{equation}
	where $K := \sqrt{1 - \frac{\xi^2}{3}}$ and $\xi = \sqrt{\frac{u_0^2}{3}} = \sqrt{\lambda^2 -1} = \sqrt{1 + \frac{\varepsilon}{2} \left(\varepsilon + \sqrt{\varepsilon^2 + 8}\right)}$.  Note that again Eq.~\eqref{eq: wave function} resembles the one obtained for relativistic ionization in \cite{Brabec, BrabecLett}. However, for ionization $\xi_I= \sqrt{1 + \frac{\varepsilon}{2} \left(\varepsilon - \sqrt{\varepsilon^2 + 8}\right)}$ \cite{Brabec, BrabecLett}, differing in one sign from $\xi$ of bound-free pair production.
	
	The short-range pair production rate results after integration over the current density at the tunnel entrance $x=l$ 
	\begin{equation*}
		\label{eq: wsr general}
		\mathcal{R}_{\text{sr}} = \frac{1}{2 \pi} \int_0^{\infty} \dd p_{\perp} \ p_{\perp} j_x (\mathbf{p}_{\perp})
	\end{equation*} 
	with  $j_x (\mathbf{p}_{\perp}) = \ p_x (x<l) \left| \Psi_{\rm sr} \left(x<l, \mathbf{p}_{\perp}\right) \right|^2$ denoting the current density  \cite{Brabec, BrabecLett}.  Exploiting $\kappa \gg \frac{1}{r}$ to simplify the spinor of Eq.~\eqref{eq: wavefunction bound state short range} results in the current density
	\begin{equation}
		\label{eq: current density}
		j_x (\mathbf{p}_{\perp}) = \frac{8 \pi^2 N^2 K^2}{1+\varepsilon} \exp \left( -2 \text{Im} \left[ \int_{l}^{0} p_x \ \dd x \right] \right).
	\end{equation} 
	Compared to the current density of relativistic ionization \cite{Brabec, BrabecLett}, the integration limits are switched as the electron for pair production moves from the tunnel entrance to the bound state, while it starts in the bound state and then exits the tunnel into the continuum for relativistic tunnel ionization. 
	
	Consequently, the short-range pair production rate is given by
	\begin{equation}
		\label{Rsr_integral}
		\begin{split}
			\mathcal{R}_{\text{sr}} &= \frac{2 N^2 K^2}{1 + \varepsilon} \int \dd p_y \int \dd p_z \exp \left( -2 \text{Im} \left[ \int_{l}^{0} p_x \dd x \right] \right) \\
			& = \frac{2 N^2 K^2}{1 + \varepsilon} \int_{-\infty}^{\infty} \dd (\delta p_y) \int_{-\infty}^{\infty} \dd (\delta p_z) \\
			& \ \ \ \ \exp \Biggl( \frac{2 c^3}{3F}  \frac{\Bigl(1+ \bigl(\frac{\delta p_y}{c} \bigr)^2 + \bigl(\frac{ p_{z0} + \delta p_z}{c} \bigr)^2 - \varepsilon^2 \Bigr)^{3/2}}{-p_{z0} / c - \delta p_z  / c+ \varepsilon} \Biggr)
		\end{split}
	\end{equation}
	with $p_x = p_{xc} = -ic \sqrt{1+(\frac{p_y}{c})^2 + (\frac{p_z}{c})^2 - \left(-\varepsilon + \frac{Fx}{c^2} \right)^2 }$, where $p_y$ and $p_z$ are rewritten according to $p_y= \delta p_y$ and $p_z = p_{z0} - Fx /c + \delta p_z$. Note that the integral over $\dd x$ is solely evaluated at $x=0$ as after an expansion around $\delta p_y = 0 = \delta p_z$ the zero- and first order terms vanish, while the second order term gives an unphysical divergence.
	
	After expansion of the resulting exponent around $\delta p_y = 0$ and $\delta p_z = 0$  up to second order and further evaluation of the resulting Gaussian integrals, we find
	\begin{equation}
		\begin{split}
		\label{eq:sr rate}
		\mathcal{R}_{\text{sr}} =& \frac{2 \pi c^2 N^2}{1 + \varepsilon} \frac{1}{\sqrt{3} \xi} \sqrt{\frac{3 - \xi^2}{3 + \xi^2}} \frac{F}{F_c} \exp \left( - \frac{2 \sqrt{3} \xi^3}{\xi^2 +1 } \frac{F_c}{F} \right)
		\end{split}
	\end{equation}
	where $F_c = c^3$ is the critical field strength. It is worth noting that the argument of the exponential function results from the zero order of the expansion while the first order vanishes and the second order yields the prefactor.
	
	The short-range pair production rate in Eq.~\eqref{eq:sr rate} looks similar to the relativistic ionization rate obtained in \cite{Brabec, BrabecLett}. However, as the variable $\xi$ contains an opposite sign in both processes, the resulting rates differ significantly. In case of vanishing binding energy, i.e. $\varepsilon \rightarrow 1$, we obtain $\xi \rightarrow \sqrt{3}$ and $\xi_I \rightarrow 0$. The exponential of relativistic ionization, which determines the main dependence of the rate on the applied field strength $F$, would be close to $1$. On the other hand, for bound-free pair production the exponential still significantly reduces the pair production rate as $\exp \left( - \frac{2 \sqrt{3} \xi^3}{\xi^2 +1 } \frac{F_c}{F} \right) \rightarrow \exp \left( - \frac{9}{2} \frac{F_c}{F} \right)$.
	
	\subsection{Coulomb correction in WKB-theory}
	\label{subsec:WKB_Q}
	Up to this point, the Coulomb field of the atom was disregarded, both for the bound and the continuum state. To include the effect of the Coulomb field, a Coulomb correction needs to be introduced \cite{Brabec, BrabecLett, ImaginaryTime1, ImaginaryTime2, ImaginaryTime_QM, TunnelMPI}. To this end, the barrier length is divided into distances close to the atomic core, such that the Coulomb field can be included by a Coulomb logarithm in the wave function, and distances further away from the atomic core, where the Coulomb potential can be treated perturbatively. Both regions are linked at an arbitrary matching point $\mathbf{r} = \mathbf{r}_1$ which fulfills $a_0 \ll r_1 \ll l$ \cite{Brabec, BrabecLett, ImaginaryTime1, ImaginaryTime2, ImaginaryTime_QM, TunnelMPI} with $a_0 = 1/Z$ referring to the mean distance of the bound electron to the atomic core with nuclear charge $Z$. Note that in WKB-theory the arbitrary matching point $r_1$ needs to cancel out in the full pair production rate \cite{LandauQM, ImaginaryTime2}. 
	
	The Coulomb correction is defined as \cite{Brabec}
	\begin{equation}
		\label{eq:Coulomb correction formula}
		Q_{\text{WKB}} = \exp \left(2 i Z \int_0^{t_1} \frac{\dd t}{r(t)}\right) \left( \frac{\Phi_{1s}}{\Psi_{\rm sr}} \bigg|_{r=r_1} \right)^2 .
	\end{equation}
	Here, $\Psi_{\rm sr}$ is the wave function of the electron in the short range potential in Eq.~\eqref{eq: wavefunction bound state short range} and $\Phi_{1s}$ describes the wave function of the hydrogenlike ground state in the Coulomb-Dirac atom which is given by
	\begin{equation}
	\begin{split}
	\label{eq: Phi_1s}
	\Phi_{1s} (x) = g(r) \chi_s \exp \left( -i E_{1s} t \right).
	\end{split}
	\end{equation}
	Here, $g(r) =C_{1s} (2Zr)^{\varepsilon -1} \exp \left( -Zr \right)$ describes the radial component with  $C_{1s} = \left(\frac{Z^3}{\pi} \frac{1 + \varepsilon}{\Gamma (1+ 2 \varepsilon)} \right)^{1/2}$ while
	\begin{equation}
		\label{spinor1}
		\chi_{+1/2}= \left(\begin{array}{c} 1 \\ 0 \\ ic \frac{1- \varepsilon}{Z} \cos \vartheta \\ ic \frac{1- \varepsilon}{Z} \sin \vartheta \ \text{e}^{i \varphi} \end{array}\right)
	\end{equation}
	and 
	\begin{equation}
		\label{spinor2}
		\chi_{-1/2}= \left(\begin{array}{c}0 \\ 1 \\ ic \frac{1- \varepsilon}{Z} \sin \vartheta \ \text{e}^{-i \varphi} \\ ic \frac{1- \varepsilon}{Z} \cos \vartheta \end{array}\right)
	\end{equation}
	denote the two possible spinors.
	
	 The Coulomb logarithm is included in the wave functions ratio. With the resulting trajectory $r(t)$ from \eqref{trajectory}, the integral in Eq.~\eqref{eq:Coulomb correction formula} is determined to
	\begin{equation}
		\begin{split}
		\label{eq:Q1}
		Q_{\text{WKB}} = & \exp \left(- \frac{6 Z}{c} \varphi_1\right) \left( \frac{\sin \left(\varphi_0 + \varphi_1\right)}{\sin \left(\varphi_0 - \varphi_1\right)} \right)^{2 \delta} \left( \frac{\Psi_c}{\Psi_{sr}} \bigg|_{r=r_1} \right)^2 ,
		\end{split}
	\end{equation}
	where $\delta = Z \varepsilon \left(1- \varepsilon^2 \right)^{-1/2} /c = \varepsilon$, $\varphi_0 = \arcsin \left(u_0 /3 \right)$,  $\varphi_1 = \arcsin \left(u_1 /3 \right)$ and $u_1$ results from the implicit formula $u_1 = \sqrt{u_0^2 - \frac{2F \lambda}{c^2} r_1 \left(1 - \frac{u_1^2}{9}\right)^{-1/2}}$. In contrast to the ionization process, the sign in the exponential of the Coulomb correction has changed due to the reversed initial and final state. Using trigonometric relations for the $\text{sin}$-function, the explicit dependence of the Coulomb correction on the matching point $r_1$ in fact cancels out in the wave functions ratio, yielding
	\begin{equation}
		\begin{split}
		\label{eq:Q2}
		 Q&_{\text{WKB}}= \frac{1 + \varepsilon}{4 \pi \Gamma \left(1 + 2 \varepsilon\right) N^2} \exp \left(- \frac{6 Z}{c} \varphi_1\right) \\
		& \times \left( \frac{Z c^2}{F \lambda} \sqrt{1-\frac{u_1^2}{9}} \left( u_0 \sqrt{1-\frac{u_1^2}{9}}+ u_1 \sqrt{1-\frac{u_0^2}{9}} \right)^{2} \right)^{2 \varepsilon}.
		\end{split}
	\end{equation}
	However, the resulting Coulomb correction still implicitly depends on the matching point $r_1$ through $u_1$. Physically, it is unreasonable to obtain a Coulomb correction factor which still depends on the arbitrary matching point. It is therefore necessary to make appropriate approximations to cancel out such a dependence. To this end we set $u_1 \approx u_0$, similarly to the calculation of the Coulomb correction for relativistic ionization \cite{Brabec, BrabecLett} and obtain for the Coulomb correction factor
	\begin{equation}
		\begin{split}
		\label{eq:Q}
		Q_{\text{WKB}} = & \frac{1 + \varepsilon}{4 \pi \Gamma \left( 1 + 2 \varepsilon \right)} \frac{1}{N^2}  \exp \left( -\frac{6 Z}{c} \arcsin \left(\frac{\xi}{\sqrt{3}}\right)\right) \\
		& \times \left( \frac{4 \xi^3 \left(3 - \xi^2\right)^2}{\sqrt{3} \left(1 + \xi^2\right)} \frac{F_c}{F}\right)^{2 \varepsilon}.
	\end{split}
	\end{equation} 
	The influence of the long-range Coulomb potential on the pair production rate is included by multiplication of the short-range rate  in Eq.~\eqref{eq:sr rate} with the Coulomb correction factor in Eq.~\eqref{eq:Q}, leading to
	\begin{equation}
		\label{eq: wpp}
		\begin{split}
			\mathcal{R}_{\text{WKB}} = & \frac{c^2}{2 \sqrt{3} \xi} \sqrt{\frac{3 - \xi^2}{3 + \xi^2}} \frac{1}{\Gamma \left(1 + 2 \varepsilon\right)} \cdot \left(\frac{F_c}{F}\right)^{2 \varepsilon -1} \\
			& \times \left(\frac{4 \xi^3 \left(3 - \xi^2\right)^2}{\sqrt{3} \left(1 + \xi^2\right)}\right)^{2 \varepsilon} \\
			& \times \exp \left(-\frac{6Z}{c} \arcsin\left(\frac{\xi}{\sqrt{3}}\right) - \frac{2 \sqrt{3} \xi^3}{1 + \xi^2} \cdot \frac{F_c}{F}\right).
		\end{split}
	\end{equation}
	From a conceptional perspective, it is worth mentioning that a similar calculation can be carried out from the positron's (i.e. hole's) point of view which starts in a bound $1s$ state and moves into the negative energy continuum. Such a calculation yields the same pair production rate as Eq.~\eqref{eq: wpp}.
	
	Concluding this section, we also note that -- during the production process -- both particles are subject to the full nuclear charge $Z$. This is intuitively visualized in the Dirac sea picture, which describes the process as the transition of an electron from the negative-energy continuum into a bound atomic state (see also Fig.~\ref{fig:schematische-darstellung-gfpp}). In this context we mention, for comparison, that in calculations of bound-free electron-positron pair production by a single high-energy photon in a nuclear Coulomb field both particles are described by Coulomb states, which account for their interaction with the nuclear $Z/r$ potential to all orders, while the interaction with the photon field is treated perturbatively \cite{Agger, Sommerfeldt}.
	
	\section{Bound-free pair production in GM-gauge SFA}
	\label{sec:SFA}
	
	Alternatively to the calculation in WKB tunneling-theory, the electron-positron bound-free pair production can also be described by an $S$-matrix formalism using the SFA in G\"oppert-Mayer (GM) gauge. A similar consideration was done in \cite{Alex} for relativistic ionization which we use as orientation and adjust it to the bound-free pair production, such that the derivation differs in several signs. 
	
	\subsection{Pair production rate in GM-gauge SFA}
	\label{subsec:SFA_rate}
	
	We start our consideration with the transition amplitude of the SFA as given in Eq.~\eqref{S_general}. 
	Note that, in contrast to the derivation in WKB tunneling theory, we consider the bound state $\Phi_{1s}$ of Eq.~\eqref{eq: Phi_1s} in a Coulomb potential and not in a short range potential. 
	
	Furthermore, compared to the consideration in Sec.~\ref{sec:Preliminary_consideration} we consider a constant crossed field using the GM-gauge such that the interaction $V^\mu$ is given by the four potential
	\begin{equation}
		\label{göppertmayergauge}
		A^{\mu}_{\rm GM} (x) = \left(- \mathbf{F} \cdot \mathbf{r}, \ - \mathbf{e}_z \left(\mathbf{F} \cdot\mathbf{r}  \right)\right).
	\end{equation}
	In the radiation gauge, the four potential of a CCF is described by $A^{\mu} = \tilde{a}^{\mu} \varphi_k$ with $\tilde{a}^{\mu} = \left( 0, \ -a , \ 0, \ 0 \right)$ and the linear phase $\varphi_k = k \cdot x$. Similar to the consideration of relativistic ionization in \cite{Alex} it is useful to introduce the auxiliary wave four-vector $k^{\mu} = \frac{\omega}{c} \left(1, \ \mathbf{e}_z \right)$ with a frequency $\omega$ which should cancel out in the final result.
	
	The corresponding amplitude for bound-free pair production thus reads 
	\begin{equation*}
		\mathcal{S}_{\rm SFA}^{\rm (GM)} = \frac{i}{c} \int d^4x\,\overline{\Phi}_{1s} \slashed{A}_{\rm GM} \Psi_{p,s}^{(\rm GM)}\,.
	\end{equation*}
	The Volkov state for the created positron is given by
	\begin{equation}
		\label{Volkov-state}
		\begin{split}
			\Psi_{p,s}^{(\rm GM)} = & \sqrt{\frac{c}{p_0}} \left(1 + \frac{\slashed{k} \slashed{A}}{2c(k \cdot p)} \right) v_{s^+} \exp \left(iS^{(+)}\right) \\
			& \times \exp \left(-i (\tilde{a} \cdot x) \varphi_k / c \right)
		\end{split}
	\end{equation}
	with the action 
	\begin{equation}
		\label{action}
		S^{(+)} = \left(p \cdot x \right) + \frac{1}{c (k \cdot p)} \left( \frac{(p \cdot \tilde{a})}{2} \varphi_k^2 + \frac{a^2}{6c} \varphi_k^3 \right)
	\end{equation}
	$v_{s^+}$ denotes the positron's  free Dirac spinor, while $p^{\mu} = (p_0, \mathbf{p})$ with $p_0 = \varepsilon_p / c$ describes the asymptotic positron momentum. The second exponential function in \eqref{Volkov-state} results from the transformation from the radiation to the G\"oppert-Mayer gauge.
	
	Similar to \cite{Alex}, we follow the treatment of ionization in \cite{Nikishov1, Nikishov2} by expressing the $\varphi_k$-dependent part of the integrand in \eqref{S_general} by a Fourier integral which yields
	\begin{widetext}
		\begin{equation}
			\label{Fourierintegral1}
			\left(1+ \frac{\slashed{k} \slashed{A}}{2c (k \cdot p)} \right) \exp \left(i \frac{(p \cdot \tilde{a})}{2c (k \cdot p)} \varphi_k^2 + i \frac{a^2}{6c^2 (k \cdot p)} \varphi_k^3 - i \frac{(\tilde{a} \cdot x)}{c} \varphi_k \right) = \int_{- \infty}^{\infty} \dd s \ \text{e}^{-is \varphi_k} \left( \mathcal{A}(s) - i \frac{\slashed{k} \slashed{\tilde{a}}}{2c (k \cdot p)} \mathcal{A}' (s) \right)
		\end{equation}
			with
		\begin{equation}
			\label{Fourierintegral2}
				\mathcal{A} (s) = \frac{1}{2 \pi} \int_{-\infty}^{\infty} \dd \varphi_k \text{e}^{i \varphi_k s} \exp \bigg(  i \frac{(p \cdot \tilde{a})}{2c (k \cdot p)} \varphi_k^2 + i \frac{a^2}{6c^2 (k \cdot p)} \varphi_k^3 - i \frac{(\tilde{a} \cdot x)}{c} \varphi_k \bigg),
		\end{equation}
	\end{widetext}
	where $\mathcal{A}'(s)$ denotes the derivative of $\mathcal{A}(s)$ with respect to $s$.  With this expression, the integral in \eqref{S_general} can be solved by first performing the integral in $t$ and then exploiting the resulting factor of $2 \pi \delta \left(\varepsilon_p + E_{1s} - s \omega \right)$ to calculate the integral in $s$. By further introducing the vector $\mathbf{q} = \mathbf{p} - s \mathbf{k} + \frac{a}{c} \varphi_k \mathbf{e}_x$, the variables $\epsilon = \frac{p \cdot \tilde{a}}{c \left(k \cdot p\right)}$, $\beta = \frac{a^2}{8 c^2 \left(k \cdot p\right)}$ and $y = \left(4 \beta \right)^{2/3} \big[ \frac{s}{4 \beta} - \big(\frac{\epsilon}{8 \beta}\big)^2 \big]$ with $s=(\varepsilon_p + E_{1s}) / \omega$ and substituting $z = \left(4 \beta\right)^{1/3} \big(\varphi_k+ \frac{\epsilon}{8 \beta}\big)$, the transition amplitude results in
	\begin{equation}
		\label{transitionamplitude}
		\begin{split}
		\mathcal{S}_{\rm SFA}^{(\rm GM)} = &  \frac{i}{\omega} \sqrt{\frac{c}{p_0}} \left(4 \beta \right)^{-1/3} \int_{-\infty}^{\infty} \dd z \int \dd^3 \mathbf{r} \ g(r) \text{e}^{-i \mathbf{q} \cdot \mathbf{r}} \bar{\chi}_{s^-} \\
		& \times \slashed{A}_G \left(1 + \frac{\slashed{k} \slashed{\tilde{a}}}{2 c \left(k \cdot p\right)} \varphi_k \right) v_{s^+} \\
		& \times \exp \left[i \left(yz + \frac{z^3}{3}\right)\right] \exp \left[i \frac{8 \beta}{3} \left(\frac{\epsilon}{8 \beta}\right)^3 - i \frac{\epsilon s}{8 \beta} \right].
	\end{split}
	\end{equation}
	From the transition amplitude \eqref{transitionamplitude} the differential pair production rate results after building the absolute square of $\mathcal{S}_{\rm SFA}^{(\rm GM)}$ and summation over the spins of both electron and positron. We use the treatment of bound-free pair production \cite{MVG-PRA2004, MVG-PRL} and relativistic ionization \cite{Reiss_JOpt} in a circularly polarized laser field in the radiation gauge as orientation and adapt the calculation to the G\"oppert-Mayer gauge \cite{Alex}. By considering the G\"oppert-Mayer gauge the calculation of the traces over the Dirac-matrices simplifies significantly as the four-products $A^2_G (x) = A_G(x) \cdot A(x) = A_G(x) \cdot k = A(x) \cdot k = k^2 = A_G(x) \cdot A_G(x')=0$ vanish, such that the absolute square of the transition amplitude reads
	\begin{equation}
		\label{absolutesquareS}
		\begin{split}
			\sum_{s^-,s^+} & \left|\mathcal{S}_{\text{SFA}}^{(\rm GM)} \right|^2 = \frac{2F^2}{p_0 \omega^2} (4 \beta)^{-2/3} \left(p_0 - p_z \right)  \\
			& \times \int \dd z \int \dd z' \exp \left[-i (yz+z^3 /3) \right]  \exp \left[i (yz'+z'^3 /3) \right] \\
			& \times \int \dd^3 \mathbf{r} \int \dd^3 \mathbf{r}' \text{e}^{i \mathbf{q} \cdot \mathbf{r}} \text{e}^{- i \mathbf{q}' \cdot \mathbf{r}'} g(r) g(r') (c_0 + c_3) \\
			& \times r' \sin \vartheta' \cos \varphi ' r \sin \vartheta \cos \varphi
		\end{split}
	\end{equation}
	with
	\begin{equation}
		\label{c0,c3}
		\begin{split}
			c_0 &= 1 + \tau^2 \left[\cos \vartheta \cos \vartheta' + \sin \vartheta \sin \vartheta' \cos (\varphi - \varphi' ) \right], \\
			c_3 &= i \tau (- \cos \vartheta + \cos \vartheta'), \\
			\tau &= (1- \varepsilon)c/Z 
		\end{split}
	\end{equation}
	The spatial integrations over $\dd^3 \mathbf{r}$ and $\dd^3 \mathbf{r}'$ can be directly carried out as in \cite{Reiss_JOpt}, while the $\dd z$ and $\dd z'$ are solved by the saddle-point method in analogy to \cite{Alex, Gribakin}. As the physically relevant saddle point $z_0 = i \sqrt{y}$ is complex, also the components of $\mathbf{q}'$ at the saddle point are complex
	\begin{equation*}
		\begin{split}
			q'_x &= \frac{a z_0'}{c (4 \beta)^{1/3}} = i \frac{\sigma}{c} , \ \ \ q'_{\perp} = \sqrt{q_x^2 + q_y^2} = i \frac{\zeta}{c}, \\
			q' &= \sqrt{q_x^2 + q_y^2 + q_z^2} = i \frac{\varrho}{c},
		\end{split}
	\end{equation*}
	with the abbreviations
	\begin{equation}
		\label{abbreviations}
		\begin{split}
			\gamma &= p_0 - p_z, \ \ \ \varrho = \sqrt{c^4 - E_{1s}^2}, \\
			\sigma &= \sqrt{\varrho^2 + p_y^2 c^2 + (c \gamma + E_{1s})^2}, \\
			\zeta &= \sqrt{\varrho^2 + (c \gamma + E_{1s})^2}.
		\end{split}
	\end{equation}
	The pair production rate results after integration over the momentum space and dividing by the interaction time $T$, 
	\begin{equation}
		\label{generalform_rate}
		\mathcal{R}_{\text{SFA}} = \int \frac{\dd ^3 \mathbf{p}}{(2 \pi)^3} \frac{\left|\mathcal{S}_{\text{SFA}}^{(\rm GM)}  \right|^2}{T}.
	\end{equation}
 	Exploiting that $\left|\mathcal{S}_{\text{SFA}}^{(\rm GM)} \right|^2$ does not depend on $p_x$ and integrating over $\gamma$ instead of $p_z$ yields
 	\begin{widetext}
 		\begin{equation}
 			\label{rate}
 			\begin{split}
 			R_{\text{SFA}} =& \frac{F^2 C_{1s}^2}{2} (2Z)^{2 \varepsilon - 2} \int_{- \infty}^{\infty} \dd p_y \int_{0}^{\infty} \dd \gamma \frac{\gamma}{\sigma} \text{e}^{-(4/3)y^{3/2}} \big\{ \big| \tilde{S}_1 (p_y, \gamma) \big|^2 + \tau^2 \big[ \big| \tilde{S}_2 (p_y, \gamma) \big|^2 + \big| \tilde{S}_3 (p_y, \gamma) \big|^2 + \big| \tilde{S}_4 (p_y, \gamma) \big|^2  \big] \\
 			& - 2 \tau \text{Re} \big[\tilde{S}_1 (p_y, \gamma)\big]  \text{Im} \big[\tilde{S}_2 (p_y, \gamma)\big] \big\},
 			\end{split}
 		\end{equation}
 		where 
 		\begin{equation}
 			\label{rate_S}
 			\begin{split}
 				\tilde{S}_1 (p_y, \gamma) &= \sigma \left[\frac{c}{\varrho^2} \Gamma \left(\frac{\varepsilon+2}{2}\right) D^{\varepsilon+2} - \frac{c^2}{\varrho^3} \Gamma \left(\frac{\varepsilon+1}{2}\right) D^{\varepsilon+1} \right], \\
 				\tilde{S}_2 (p_y, \gamma) &= i \sigma q_z \left[-3 \frac{c^4}{\varrho^5} \Gamma \left(\frac{\varepsilon}{2} \right) D^{\varepsilon} + 3\frac{c^3}{\varrho^4} \Gamma \left(\frac{\varepsilon+1}{2}\right) D^{\varepsilon+1} - \frac{c^2}{\varrho^3} \Gamma \left(\frac{\varepsilon+2}{2}\right) D^{\varepsilon+2} \right],\\
 				\tilde{S}_3 (p_y, \gamma) &= \frac{\sigma}{\zeta} \left[\left(3 \frac{\zeta^2 c^3}{\varrho^5} - \frac{c^3}{\varrho^3}\right) \Gamma \left(\frac{\varepsilon}{2} \right) D^{\varepsilon} + \left(\frac{c^2}{\varrho^2} - 3 \frac{\zeta^2 c^2}{\varrho^4}\right) \Gamma \left(\frac{\varepsilon+1}{2}\right) D^{\varepsilon+1} - \frac{\zeta^2 c}{\varrho^3} \Gamma \left(\frac{\varepsilon+2}{2}\right) D^{\varepsilon+2} \right],\\
 				\tilde{S}_4 (p_y, \gamma) &= -i \frac{p_y c}{\zeta} \left[ \frac{c^3}{\varrho^3} \Gamma \left(\frac{\varepsilon}{2} \right) D^{\varepsilon} - \frac{c^2}{\varrho^2} \Gamma \left(\frac{\varepsilon+1}{2}\right) D^{\varepsilon+1} \right],
 			\end{split}
 		\end{equation}
 		with $D= Zc \big( \frac{2}{F \gamma \sigma} \big)^{1/2}$. 
 	\end{widetext}
 	The remaining integrals can be carried out numerically to obtain the final pair production rate. Alternatively, the pair production rate \eqref{rate} can be simplified analytically, following the approach used for the relativistic ionization rate in \cite{Alex}. To this end, we assume that the exponential term $\exp (- (4/3) y^{3/2})$ varies only slowly with respect to $\gamma$ and $p_y$. In analogy to \cite{Alex}, the $\gamma$-integration is simplified by only integrating the exponential which we expand up to second order around $$\gamma_0 = - \frac{E_{1s}}{4c} + \frac{1}{4c} \sqrt{E_{1s}^2 + 8 c^4}$$ where the exponential takes on its maximum value. As no significant contribution to the integral results from negative values of $\gamma$, we further formally extend the lower boundary of the integration to $- \infty$. Performing these simplifications results in two Gaussian integrals which can be evaluated analytically. In total the simplified bound-free pair production rate yields 
 	\begin{equation}
 		\label{rate_simplified}
 		\begin{split}
 			\mathcal{R}&_{\text{SFA}} = \frac{1}{2} \pi F^2 C_{1s}^2 (2Z)^{2 \varepsilon -2} \frac{\gamma_0}{\sigma_0} \sqrt{\frac{2F \gamma_0}{\sigma_0 h'' (\gamma_0)}} \text{e}^{-h(\gamma_0)} \\
 			&\times  \big\{ \big| \tilde{S}_1 (p_y = 0, \gamma = \gamma_0) \big|^2 \\
 			& + \tau^2 \big[ \big| \tilde{S}_2 (p_y=0, \gamma = \gamma_0) \big|^2 + \big| \tilde{S}_3 (p_y = 0, \gamma = \gamma_0) \big|^2  \big] \\
 			& -2 \tau \text{Re} \big[\tilde{S}_1 (p_y=0, \gamma = \gamma_0 )\big]  \text{Im} \big[\tilde{S}_2 (p_y=0, \gamma = \gamma_0)\big]\big\} ,
 		\end{split}
 	\end{equation}
 	where
 	\begin{equation*}
		\begin{split}
			\sigma_0 &= \sqrt{\varrho^2 + (c \gamma_0 + E_{1s})^2} \\
			h (\gamma) &= \frac{2 \sigma^3}{3 c^2 F \gamma}.
		\end{split}
 	\end{equation*}
 	We emphasize that the exponential $\exp \left(-h(\gamma_0)\right)$ of Eq.~\eqref{rate_simplified} can be rewritten such that it exactly coincides with the exponential of the WKB-rate $\mathcal{R}_{\text{WKB}}$ in Eq.~\eqref{eq: wpp}.
 	
	\subsection{Coulomb correction in GM-gauge SFA}
	\label{subsec:SFA_Q}
	For the calculation of bound-free pair production in WKB-theory, we have already pointed out that a multiplication with a Coulomb correction factor is necessary to include the effect of the disregarded long-range Coulomb potential. Similarly, in the calculation in GM-gauge SFA, the influence of the Coulomb field on the continuum state so far has not been taken into account. However, since our WKB tunneling treatment relied on a bound $1s$ state in a short-range $\delta$-potential, the necessary correction factor \eqref{eq:Q} takes the long-range Coulomb field fully into account. In contrast, our GM-gauge SFA calculation already contains the Coulomb field in the bound $\Phi_{1s}$. As a consequence, the Coulomb correction factor in Eq.~\eqref{eq:Q} overestimates the influence of the Coulomb field in the pair production rate obtained through the GM-gauge SFA. To appropriately account for the remaining Coulomb correction, we therefore need to truncate the correction factor in Eq.~\eqref{eq:Q}. To this end, we only consider the perturbative part of the Coulomb correction factor, i.e. the exponential including the integration in Eq.~\eqref{eq:Coulomb correction formula}. Similar to the consideration of relativistic ionization in \cite{Alex} a point $r_1$ with time $t_1$, up to which we take the Coulomb correction into account, needs to be estimated. Consequently the Coulomb correction reads \cite{Brabec, BrabecLett, ImaginaryTime1, ImaginaryTime2, ImaginaryTime_QM, TunnelMPI, Alex}
	\begin{equation}
		\label{eq: Q general SFA}
		Q_{\text{SFA}}= \exp \left(2iZ \int_{0}^{t_1} \frac{1}{r(t)} \dd t \right) ,
	\end{equation}
	where $t=0$ denotes the tunnel entrance. 
	
	We determine $r_1$ and the associated value for $u_1$ by the intuitive approach that at $r_1$ the Coulomb field strength and the CCF amplitude coincide \cite{Alex}. This leads to 
	\begin{equation}
		\label{r1}
		r_1= \left(1- \frac{u_1^2}{9}\right)^{1/4} \sqrt{\frac{Z}{F}}
	\end{equation}
	where $r_1 = \frac{c^2}{2F \lambda} \left(u_0^2 - u_1^2 \right) \sqrt{1- \frac{u_1^2}{9}}$ follows from the parametric form of the trajectory \eqref{trajectory}. For relativistic ionization a similar consideration was done in \cite{Alex}. Note however, that for the ionization process $r \approx x$ holds, as $r = x \sqrt{1- \frac{u_1^2}{9}}$ and $\sqrt{1- \frac{u_1^2}{9}} \rightarrow 1$, which simplifies the result for $r_1$ and $u_1$ significantly. 
	
	According to Eq.~\eqref{eq:Q1} we find for the Coulomb correction
	\begin{equation}
		\label{eq:Q SFA}
			Q_{\text{SFA}} = \exp \left(- \frac{6 Z}{c} \varphi_1\right)  \left( \frac{\sin \left(\varphi_0 + \varphi_1\right)}{\sin \left(\varphi_0 - \varphi_1\right)} \right)^{2 \delta}
	\end{equation}
	where $\delta = Z \varepsilon \left(1- \varepsilon^2 \right)^{-1/2} /c = \varepsilon$, $\varphi_0 = \arcsin \left(u_0 /3 \right)$ and $\varphi_1 = \arcsin \left(u_1 /3 \right)$ with $u_1$ resulting from Eq.~\eqref{r1}.
	
	\begin{figure}[t]  
		\vspace{-0.25cm}
		\begin{center}
			\includegraphics[width=0.48\textwidth]{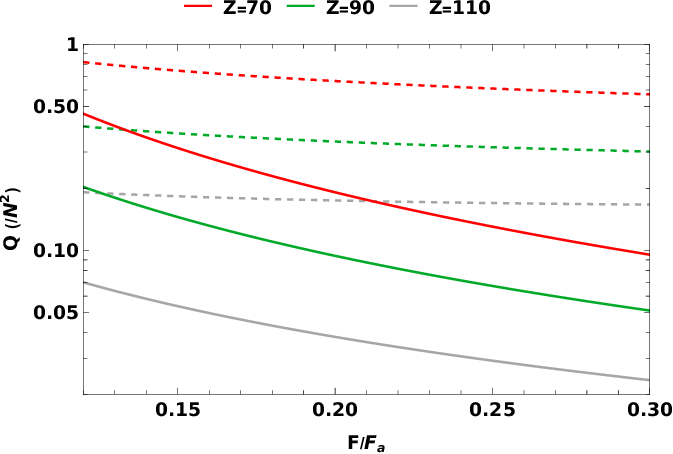}
		\end{center}
		\vspace{-0.5cm} 
		\caption[justification=justified]{Coulomb correction factor as a function of the field strength $F$ in units of the atomic field strength $F_a=Z^3$ for different nuclear charges $Z=70,  \ 90, \ 110$. The solid curves show the full correction factor from Eq.~\eqref{eq:Q} and the dashed curves depict the truncated factor in Eq.~\eqref{eq:Q SFA}.}
		\label{fig_Qcomparison}
	\end{figure}
	
	Figure \ref{fig_Qcomparison} shows the Coulomb correction factor as a function of the normalized field strength $F / F_a$ for nuclear charges $Z=70, \ 90, \ 110$. The full Coulomb correction \eqref{eq:Q} is shown by solid lines while the truncated Coulomb correction \eqref{eq:Q SFA} is depicted by dashed lines.  Note that the full Coulomb correction \eqref{eq:Q} formally still depends on the undefined normalization constant $N$ which cancels out in the expression for the pair production rate. In Figure \ref{fig_Qcomparison}, Eq.~\eqref{eq:Q} is therefore divided by $N^2$. Figure \ref{fig_Qcomparison} shows that both Coulomb correction factors \eqref{eq:Q} and \eqref{eq:Q SFA} lead to a decrease of the pair production rate, meaning that the neglect of the Coulomb potential leads to an overestimation of the actual rate. In contrast, for relativistic ionization both Coulomb corrections lead to an increase of the ionization rate \cite{Alex}. As the full Coulomb correction was needed for the pair production rate in the WKB-theory, the resulting correction factor is smaller than the correction factor for the SFA result \eqref{eq:Q SFA}. In both cases, the Coulomb correction is smallest for large nuclear charges and decreases when the applied field $F$ grows.
	
	\section{Results and Discussion}
	\label{sec:results}
	In this section we discuss the results of our two theoretical approaches for bound-free pair production in a CCF by comparing the resulting total and differential pair production rates.  In particular we discuss the field dependent exponential function of the pair production rate by introducing certain approximations which include a physically intuitive dependence on the binding potential $I_p = c^2 -E_{1s}$. 
	
	Regarding the applied field parameters, we want to emphasize that once the electron positron pair was created,  a bound electron is present that could in principle be ionized due to the CCF. For a large enough lifetime of the bound state we demand that the reciprocal of the orbital time $T_a$ needs to be larger than the ionization rate. The rate $\mathcal{R}_I$ for relativistic tunneling ionization in a CCF was obtained in \cite{Brabec} in Eq.~(53). For the nuclear charges $Z=70$, $Z=90$ and $Z=110$ the maximal field strength $F$, where $\mathcal{R}_I = 1 / T_a$, can therefore be estimated as $F=0.27 \ F_a$, $F=0.31 \ F_a$ and $F=0.37 \ F_a$ respectively. As a consequence we consider the pair production rates up to $F=0.3 \ F_a$. We point out that this upper limit may be regarded as a conservative estimate, because ionization at $F \lesssim F_a$ already proceeds in the over-barrier regime where tunneling rates are known to overestimate the ionization yield \cite{BauerMulser, TongLin, Klaiber}.
	
	\subsection{Total pair production rates}
	\label{subsec:results_rate}
	
	\begin{figure}[t]  
		\vspace{-0.25cm}
		\begin{center}
			\includegraphics[width=0.48\textwidth]{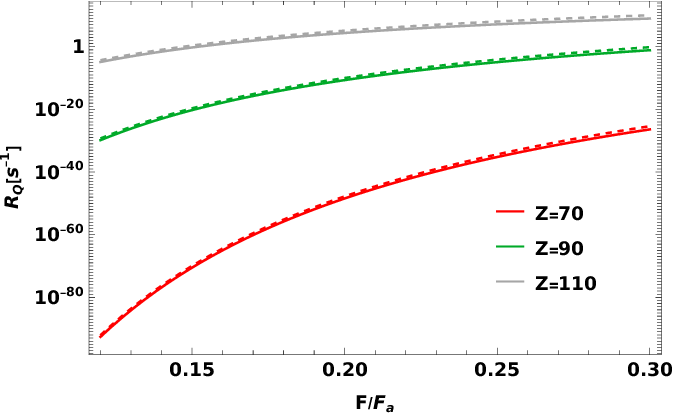}
		\end{center}
		\vspace{-0.5cm} 
		\caption[justification=justified]{Total pair production rate as a function of the field strength $F$ in units of the atomic field strength $F_a=Z^3$ for different nuclear charges $Z=70,  \ 90, \ 110$. The solid curves show the analytical expression in Eq.~\eqref{eq: wpp} and the dashed curves depict Eq.~\eqref{rate_simplified} multiplied by the truncated Coulomb correction factor \eqref{eq:Q SFA}.}
		\label{fig_ratecomparison}
	\end{figure}
	
	In Figure \ref{fig_ratecomparison} we compare both results for the bound-free pair production rate in a combined Coulomb field and a CCF. The solid lines depict the analytically obtained expression in Eq.~\eqref{eq: wpp} and the dashed curves show Eq.~\eqref{rate_S} multiplied by the Coulomb factor \eqref{eq:Q SFA} as a function of the normalized field strength $F/F_a$. For both results, the exponential dependence can be clearly seen in the course of the curves. Figure \ref{fig_ratecomparison} shows the pair production rate on a logarithmic scale which extends over about $100$ orders of magnitude. On this scale both approaches give very similar pair production rates and only differ slightly. When considering the ratio of both rates for large fields $F=0.3 \ F_a$, the WKB rate \eqref{eq: wpp} is approximately $1$ order of magnitude smaller than the SFA rate. For smaller field strengths $F=0.12 \ F_a$, the SFA rate is only larger by a factor of approximately $3$. For example for $Z=70$ the rate spans over $50$ orders of magnitude. A deviation of both results, which coincide in the general course of the curve of $1$ order of magnitude, is therefore not as significant.
	
	The main dependence of the pair production rate on the applied field $F$ lies in the exponential function which is obtained in both previously discussed approaches, while the prefactor differs in both results. This also directly follows from the fact that the exponential factor results from the zero order contribution in the WKB calculation. The previously obtained exponential function of Eq.~\eqref{eq: wpp} and Eq.~\eqref{rate_simplified} is given by
	\begin{equation}
		\label{exponent_bfpp}
		\mathcal{R}_{\text{WKB}} \sim \mathcal{R}_{\text{SFA}} \sim \exp \left(- \frac{2 \sqrt{3} \xi^3}{1 + \xi^2}  \frac{F_c}{F}\right).
	\end{equation}
	
\begin{figure}[t]  
	 	\vspace{-0.25cm}
	 	\begin{center}
	 		\includegraphics[width=0.48\textwidth]{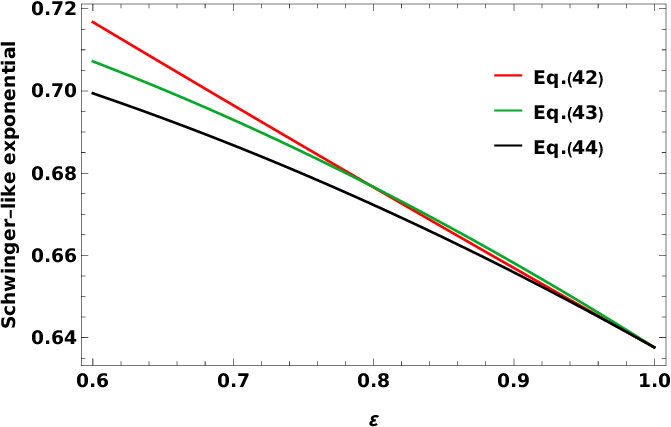}
	 	\end{center}
	 	\vspace{-0.5cm} 
	 	\caption{Comparison of the Schwinger-like exponential function for $F=0.1 \ F_c$. The exponential of bound-free pair production \eqref{exponent_bfpp} is depicted in red, the exponential  containing the physically intuitive energy barrier \eqref{exponent_energy} with $b=\frac{9}{4 \sqrt{2}}$ in green and \eqref{exponent_tunnel} which includes the tunnel length in black.} 
	 	\label{fig_exponential}
	 \end{figure}

	In Sec.~\ref{sec:Preliminary_consideration} we've presented an approximate exponential dependence, which asymptotically results for bound-free pair production in a circular polarized field in radiation gauge and through a consideration of the tunneling process. The exponential \eqref{R_RG} contains a physically intuitive dependence on the energy difference $2c^2 - I_p$, which coincides with the energy gap from the negative to the positive energy continuum lowered by the binding potential of the final bound state and has the form
	\begin{equation}
		\label{exponent_energy}
		\mathcal{R}_1 \sim \exp \left(- b \left( 2- \frac{I_p}{c^2} \right)^{3/2} \frac{F_c}{F}\right).
	\end{equation}
	In Sec.~\ref{sec:Preliminary_consideration}, $b=\frac{4 \sqrt{2}}{3}$ was obtained.	However, this consideration of the pair production process only gives an approximate dependence of the rate on the applied field. To ensure that the exponential shows a good resemblance to Eq.~\eqref{exponent_bfpp}, $b$ needs to be adjusted such that the two exponentials coincide for $I_p=0$. For $b=\frac{9}{4 \sqrt{2}}$ a good approximation for the actual exponent in \eqref{exponent_bfpp} which also includes the expected reduction of the energy barrier to $2c^2 -I_p$ (see also Fig.~\ref{fig_exponential}) is achieved. We note moreover that in the formal limit $\varepsilon\to 0$ (i.e., $I_p\to c^2$), our expression closely resembles the bound-free transition rate $\sim\exp\big(-\frac{\pi}{2}\frac{F_c}{F}\big)$ obtained in a constant electric field in \cite{ImaginaryTime1} since $\frac{9}{4\sqrt{2}} \approx \frac{\pi}{2}$ (see Eq.~(5') therein).
	 
	 A similar term can also be obtained by a Taylor expansion around vanishing binding energy $I_p = 0$ or correspondingly $\xi = \sqrt{3}$ of the exponential's argument in \eqref{exponent_bfpp}. This yields
 	\begin{equation}
 		\label{exponent_tunnel}
 			\mathcal{R}_2 \sim \exp \left( - \frac{4 }{3} \frac{F_c}{F} \left( \frac{9}{4}- \frac{I_p}{c^2} \right)^{3/2} \right).
 	\end{equation}
	Equation \eqref{exponent_tunnel} includes a reduction of $9/4 \ c^2$ by the binding potential instead of a reduction of the energy gap between the negative and positive continuum. $\frac{9c^2}{4F}$ exactly coincides with the tunnel length $l= \frac{u_0^2}{2F \lambda} c^2$ for $I_p=0$.  Due to the magnetic field, the tunnel is larger than expected from the energy difference of $2c^2$. Through the Taylor expansion we can make a connection of the pair production exponential to the actual tunnel length between negative and positive energy continuum.
	
	Figure~\ref{fig_exponential} shows that both physically motivated exponential functions \eqref{exponent_energy} with $b=\frac{9}{4 \sqrt{2}}$ and \eqref{exponent_tunnel} lead to a good resemblance with the exponential of bound-free pair production in Eq.~\eqref{exponent_bfpp}. 
	
	The characteristic dependence of the rate's exponential implies that the atomic $1s$ state gives the dominant contribution to the bound-free pair production, because in this state the electron is most tightly bound, which leads to the largest reduction of the energy gap. Let us estimate, based on the obtained exponential dependence, the relative contribution from the $2s$ state. In this case, the electron experiences a smaller Coulomb field (which scales like $1/n^3$ with the principal quantum number $n$). Therefore, we need to apply a correspondingly smaller external field strength in order to ensure that the basic physical properties of the bound $2s$ state are preserved. Taking $F\approx 0.04 \, F_a$ and $Z=70$ ($Z=110$), the exponential rate dependency predicts that the contribution from the $2s$ state is smaller by a factor of order $10^{-25}$ ($10^{-18}$) than the contribution from the $1s$ state. To further corroborate this estimate by taking the impact of the pre-exponential factor into account, we have calculated with the method of Sec.~\ref{sec:SFA} the bound-free pair production when the electron is created in the $2s$ state. The ratio of the $2s$ to the $1s$ contribution to the production rate is found as $10^{-24}$ ($10^{-16}$) for the above parameters $F\approx 0.04 \, F_a$ and $Z=70$ ($Z=110$).
	We may thus conclude that the $1s$ state is by far of the highest relevance for bound-free pair production in the considered strong-field regime. For comparison, we note that the contribution from the $2s$ state to bound-free pair production by a single high-energy photon \cite{Agger} or by several photons from a low-intensity ($\eta\ll 1$) laser wave \cite{Deneke} scales approximately with $1/n^3$ and, thus, amounts to about 13\%. An additional contribution of a few percent results when also the $2p$ states are taken into account.
	
	\subsection{Momentum distribution of emitted positrons}
	\label{subsec:results_p}
	So far, we have discussed the total pair production rate for the tunneling consideration in Sec.~\ref{sec:WKB} and the SFA calculation in Sec.~\ref{sec:SFA}. In both calculations, the total pair production rate is obtained after performing a Taylor expansion of the exponential's argument around $\delta p_y = 0$ and $\delta p_z = 0$ up to second order followed by an integration over the momenta (see below Eq.~\eqref{Rsr_integral}). In the following we consider the differential pair production rates $\frac{\dd R}{\dd ( \delta p_y )}$ and $\frac{\dd R}{\dd ( \delta p_z )}$ which result when omitting the integral over one momentum $ \delta p_y$ or $ \delta p_z$ respectively.
	
	\begin{figure}
		\centering
		\vspace{-0.25cm}
		\includegraphics[width=0.48\textwidth]{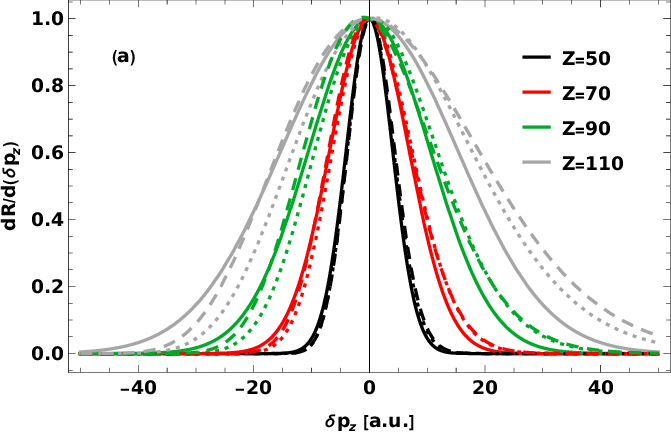}
		\vspace{0.15cm}
		\includegraphics[width=0.48\textwidth]{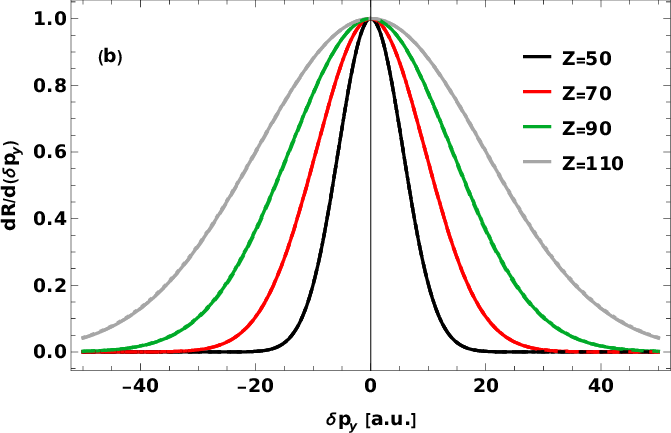}
		\vspace{-0.5cm}
		\caption{Normalized differential pair-production rates (a) $\frac{\dd R}{\dd ( \delta p_z )}$ and (b) $\frac{\dd R}{\dd ( \delta p_y )}$ for nuclear charges $Z=50, \ 70, \ 90, \ 110$ at $F=0.2 \ F_a$. The solid curves show the Gaussian distributions which result after Taylor expansion of the exponential in Eqs.~\eqref{Rsr_integral} and \eqref{rate} around $\delta p_y = 0$ and $\delta p_z = 0$. The dotted lines show the differential pair production rate \eqref{Rsr_integral} of the calculation in the WKB-theory, where one integral over the momentum has been carried out numerically. The dashed lines show the numerically resulting differential pair production rates in the SFA in \eqref{rate} where the momentum dependent prefactor is included in the integration. The maximum values at the centers of the presented Gaussian distributions amount to $\frac{\dd R}{\dd ( \delta p_y )} = \frac{\dd R}{\dd ( \delta p_z )}= 6.4 \times 10^{2} \, \text{a.u.}, \ 8.5 \times 10^{-13} \, \text{a.u.}, \ 2.2 \times 10^{-50} \, \text{a.u.} \ \text{and} \ 5.3 \times 10^{-176} \, \text{a.u.}$ for $Z = 110, \ 90, \ 70 \ \text{and} \ 50$, respectively. }
		\label{fig-momentum}
	\end{figure}
	
	Figure \ref{fig-momentum} shows the differential pair production rate $\frac{\dd R}{\dd ( \delta p_z )}$ in Fig.~\ref{fig-momentum}(a) and $\frac{\dd R}{\dd ( \delta p_y )}$ in Fig.~\ref{fig-momentum}(b) for different nuclear charges $Z$ at $F=0.2 \ F_a$. The rates shown in Fig.~\ref{fig-momentum}  are normalized to a height of $1$ to facilitate their comparison.  The resulting Gaussian distributions when performing the Taylor expansion are represented by solid lines. Here, both the approach by a tunneling consideration in the WKB-approximation and the SFA-approach yield the same momentum distribution. We recall that in the SFA-approach, the Gaussian distribution results after performing the Taylor expansion of the exponential while setting $\gamma = \gamma_0$ and $p_y =0$ in the momentum dependent prefactor of Eq.~\eqref{rate}. In the WKB-approach the prefactor is independent of the momenta such that it suffices to expand the exponential in Eq.~\eqref{Rsr_integral}. The dotted lines show the numerically calculated result of the momentum distribution obtained in the WKB-tunneling calculation in Sec.~\ref{sec:WKB}. The dashed curves show the exact differential rate obtained in the SFA in Sec.~\ref{sec:SFA}. While only the exponential function in Eq.~\eqref{Rsr_integral} needs to be integrated in the WKB-approach, the momentum dependent prefactor of Eq.~\eqref{rate} in the SFA-calculation is included in the numerically evaluated integration.
	
	The $\delta p_y$ distribution shows almost no difference between the approximated Gaussian distribution and the exact results from the WKB- and SFA- calculations. For the $\delta p_z$ distribution on the other hand, both differential rates $\frac{\dd R}{\dd ( \delta p_z )}$ which result after the numerical integration show a shift towards positive $\delta p_z$ compared to the approximated Gaussian distribution. When integrating over $\dd (\delta p_z)$, the shift approximately cancels out such that the previously performed approximations to obtain the total rate of bound-free pair production are justified. Even though our previously presented different methods yield slightly different differential pair production rates with respect to $\delta p_z$, it is worth noting that the difference between both results stems from the prefactor. In the calculation based on the WKB-theory, the prefactor does not depend on the momenta $p_y$ and $p_z$, while the SFA-calculation yields a momentum-dependent prefactor which is included in the numerical integration. If we omit the prefactor of Eq.~\eqref{rate}, by setting $p_y=0$ and $\gamma = \gamma_0$ and only calculate the integral over the exponential, the same differential rate results from our two methods.
	
	At this point it is worth mentioning that the rate ~\eqref{rate} in the SFA-calculation shows a Gaussian distribution around $\gamma = \gamma_0$ after having performed the Taylor expansion. The distribution is depicted around $\delta p_z = 0$ to facilitate the comparison to the distribution of the WKB-calculation in \eqref{Rsr_integral}. For comparison we consider the resulting $\gamma_0$ of both equations for $I_p / c^2 \ll 1$. In the SFA-calculation $\gamma_0$ results as
	\begin{equation}
	\label{gamma0_appr_SFA}
		\gamma_0 =   - \frac{E_{1s}}{4c} + \frac{1}{4c} \sqrt{E_{1s}^2 + 8 c^4} \approx \frac{c}{2} + \frac{I_p}{6c}.
	\end{equation}
	In the WKB-calculation, the corresponding value for $\gamma_0$ results from the momenta $p_z = -p_{z0} + \frac{1}{c}F x$ and $p_0=\sqrt{c^2 + p^2}$ of the created positron. Note that $p_z$ of the positron is defined with an additional minus sign compared to the momentum of the electron which we have considered in Sec.~\ref{sec:WKB}. At the tunnel exit $x=l= \frac{u_0^2 c^2}{2F \lambda}$, the momentum in $x$-direction vanishes such that $\gamma$ for $I_p / c^2 \ll 1$ reads
	\begin{equation}
		\label{gamma0_appr_WKB}
		\gamma (x=l) = c \sqrt{1 + p_z^2/c^2} - p_z \approx \frac{c}{2} + \frac{I_p}{6c},
	\end{equation}
	which coincides with $\gamma_0$ of the SFA-calculation. As we consider the birth process of the positron to be completed at the tunnel exit, the largest contribution to the pair production rate results at this point. 
	
	Fig.~\ref{fig-momentum} further shows that the positron momentum distribution is the broader, the larger the nuclear charge $Z$ is. Both for the WKB-tunneling calculation in Sec.~\ref{sec:WKB} and the calculation in SFA in Sec.~\ref{sec:SFA}, the same approximate Gaussian distribution is obtained. For $I_p / c^2 \ll 1$, the standard deviations $\sigma_y$ and $\sigma_z$ of both distributions approximately yield
	\begin{equation}
		\label{sigma_y}
		\sigma_y = \sqrt{\frac{F}{2 \sqrt{3} c \xi }} \approx \sqrt{\frac{F}{6c}} \left( 1 + \frac{2}{9} \frac{I_p}{c^2}\right)
	\end{equation}
	and
		\begin{equation}
		\label{sigma_z}
		\sigma_z = \sqrt{\frac{F \sqrt{3}}{2c \xi \left(3+\xi^2\right)}} \approx \sqrt{\frac{F}{12c}} \left( 1 + \frac{4}{9} \frac{I_p}{c^2}\right).
	\end{equation}
	The approximate standard deviations show that the width of the Gaussian distributions grows linearly with the ionization potential $I_p$ and that the width of the $p_z$-distribution is smaller than of the $p_y$-distribution. Note that even though the approximate expressions in Eqs.~\eqref{sigma_y} and ~\eqref{sigma_z} rely on the assumption that $I_p / c^2 \ll 1$, they yield good agreement with the actual width for the considered nuclear charges $Z=50 , \ 70, \ 90 \ \text{and} \ 110$.
	
	\section{Conclusion}
	\label{sec:conclusion}
	Bound-free electron-positron pair production in combined nuclear Coulomb
	and constant crossed electromagnetic fields has been studied. The electron is
	created in the $1s$ ground state of the resulting hydrogen-like ion. Two different
	approaches have been applied: (i) a WKB  tunneling theory and (ii) an SFA treatment
	in G\"oppert Mayer gauge, each incorporating an appropriate Coulomb correction factor.
	In comparison with the WKB approach, which relies on a bound state in a short-range
	potential, the SFA consideration involves a truncated version of the Coulomb correction
	since the bound state entering here already contains the long-range Coulomb interaction.
	Both methods have successfully been applied before to calculate the related process of
	relativistic strong-field ionization \cite{Brabec, BrabecLett, ImaginaryTime1, ImaginaryTime2, ImaginaryTime_QM, TunnelMPI}.
	
	Closed-form analytical expressions for the bound-free production rate have been obtained
	from both approaches in Eqs.~\eqref{eq: wpp} and \eqref{rate_simplified}, respectively. The rate is found to be very sensitive to the nuclear charge $Z$ and the external field strength $F$, depending on both parameters in a nonperturbative way. The main field dependence is of exponential form, resembling the
	Schwinger rate for free pair production in a constant electric field. However, also the atomic
	binding potential enters into the exponent and softens the exponential rate suppression. This
	'intrinsic assistance' can be explained intuitively by considering either the correspondingly
	reduced energy gap or the tunneling length of bound-free pair production.
	
	The results of our WKB and SFA calculations show good mutual agreement in a wide range of
	nuclear charges and applied field strengths. This holds for both the total production rates and the
	momentum distributions of the created positron. The latter possess an approximately Gaussian
	shape and become broader, when the nuclear charge or the external field is increased. 
	
	The considered process could, in principle, be realized experimentally by bringing an extremely relativistic beam of bare heavy ions into collision with a high-intensity optical or infrared laser wave. When the bound electron after its creation is quickly ionized in the presence of the strong external field, a free electron-positron pair results in the end. The combined process of bound-free pair production followed by strong-field ionization may therefore be considered as a two-step channel for free-free pair production. From the perspective of collision physics, it resembles a resonant scattering process \cite{Schulz}, where the resonance is associated with the intermediate bound state.
	
	\begin{acknowledgements}
		This work has been funded by the Deutsche For\-schungsgemeinschaft (DFG) 
		under Grant No.~392856280 within the Research Unit FOR 2783/2.
	\end{acknowledgements}


\begin{thebibliography}{33}
		\bibitem{Sauter}
		F.~Sauter, Über das Verhalten eines Elektrons im homogenen elektrischen Feld nach der relativistischen Theorie
		Diracs, Z. Phys. \textbf{69}, 742 (1931).
		
		\bibitem{Schwinger} J. S. Schwinger, On Gauge Invariance and Vacuum Polarization, Phys. Rev. {\bf 82}, 664 (1951).
		
		\bibitem{Ritus-Review}
		V.~I.~Ritus, Quantum effects of the interaction of elementary particles with an intense electromagnetic field, J. Sov. Laser Res.\textbf{6}, 497 (1985).
		
		\bibitem{Review1} F. Ehlotzky, K. Krajewska, and J. Z. Kami\'nski, Fundamental processes of quantum electrodynamics in laser fields of relativistic power, Rep. Prog. Phys. {\bf 72}, 046401 (2009).
		
		\bibitem{Review2} R. Ruffini, G. Vereshchagin, and S.-S. Xue, Electron–positron pairs in physics and astrophysics: From heavy nuclei to black holes, Phys. Rep. {\bf 487}, 1 (2010).
		
		\bibitem{Review3} A. Di Piazza, C. M\"uller, K. Z. Hatsagortsyan, 
		and C. H. Keitel, Extremely high-intensity laser interactions with fundamental quantum systems, Rev. Mod. Phys. {\bf 84}, 1177 (2012).
		
		\bibitem{Review4} A. Fedotov, A. Ilderton, F. Karbstein, B. King, 
		D. Seipt, H. Taya, and G. Torgrimsson, Advances in QED with intense background fields, Phys. Rep. {\bf 1010}, 1 (2023). 
		
		\bibitem{ELI} I. C. E. Turcu et al., High field physics and QED experiments at ELI-NP, Rom. Rep. Phys. {\bf 68}, S145 (2016); see also \url{https://eli-laser.eu}.
		
		\bibitem{CoReLS} J. W. Yoon et al., Realization of laser intensity over $10^{23} \ \text{W/cm}^2$, Optica \textbf{8}, 630 (2021);
		see also \url{https://corels.ibs.re.kr}.
		
		\bibitem{FACET} S. Meuren, E-320 Collaboration at FACET-II, 
		\url{https://facet.slac.stanford.edu}.
		
		\bibitem{CALA} F. C. Salgad et al., Towards pair production in the non-perturbative regime, New J. Phys. {\bf 23}, 105002 (2021).
		
		\bibitem{LUXE} H. Abramowicz et al., Conceptual design report for the LUXE experiment, Eur. Phys. J. Spec. Top. 
		{\bf 230}, 2445 (2021); see also \url{http://www.hibef.eu}.
		
		\bibitem{Gemini} C. H. Keitel {\it et al.}, Photo-induced pair production and strong field QED on Gemini, arXiv:2103.06059.
		
		\bibitem{Reiss}
		H.~R.~Reiss, Absorption of Light by Light, J. Math. Phys. \textbf{3}, 59 (1962).
		
		\bibitem{Nikishov-Ritus}
		A.~I.~Nikishov and V.~I.~Ritus, Quantum processes in the field of a plane electromagnetic wave and in a constant field I, Zh. Eksp. Teor. Fiz. \textbf{46}, 776 (1963)
		[Sov. Phys. JETP \textbf{19}, 529 (1964)].
		
		\bibitem{Yakovlev} V. Yakovlev, Electric and magnetic properties of a semiconductor in the field of a strong electromagnetic wave, Zh. Eksp. Teor. Fiz. {\bf 49}, 318 (1965)
		[Sov. Phys. JETP {\bf 22}, 223 (1966)].
		
		\bibitem{Ritus-trident} V. I. Ritus, Vacuum polarization correction to elastic electron and muon scattering in an intense field and pair electro- and muoproduction, Nucl. Phys. B {\bf 44}, 236 (1972).
		
		\bibitem{Narozhny-Nikishov} N. B. Narozhnyi and A. I. Nikishov, Electron-positron pair production by a Coulomb center located in a constant field , Zh. Eksp. Teor. Fiz. {\bf 63}, 1125 (1972) [Sov. Phys. JETP {\bf 36}, 598 (1973)].
		
		\bibitem{SLAC} D.~L.~Burke et al., 
		Positron Production in Multiphoton Light-by-Light Scattering, 
		Phys. Rev. Lett. \textbf{79}, 1626 (1997).
		
		\bibitem{Mittleman} M. H. Mittleman, Multiphoton pair creation, Phys. Rev. A {\bf 35}, 4624 (1987).
		
		\bibitem{Avetissian} H. K. Avetissian, A. K. Avetissian, G. F. Mkrtchian, 
		and K. V. Sedrakian, X-ray free electron laser for electron–positron pair production on the nuclei, Nucl. Instrum. Methods Phys. Res. A {\bf 507}, 582 (2003).
		
		\bibitem{MVG-PRA2003} C. M\"uller, A. B. Voitkiv, and N. Gr\"un, Differential rates for multiphoton pair production by an ultrarelativistic nucleus colliding with an intense laser beam, 
		Phys. Rev. A {\bf 67}, 063407 (2003).
		
		\bibitem{Sieczka} P. Sieczka, K. Krajewska, J. Z. Kami\'nski, P. Panek, 
		and F. Ehlotzky, Electron-positron pair creation by powerful laser-ion impact, Phys. Rev. A {\bf 73}, 053409 (2006).
		
		\bibitem{Krajewska-PRA2006} J. Z. Kami\'nski, K. Krajewska, and F. Ehlotzky, Monte Carlo analysis of electron-positron pair creation by powerful laser-ion impact, 
		Phys. Rev. A {\bf 74}, 033402 (2006).
		
		\bibitem{Milstein} A. I. Milstein,  C. M\"uller, K. Z. Hatsagortsyan, U. D. Jentschura, and C. H. Keitel, Polarization-operator approach to electron-positron pair production in combined laser and Coulomb fields, Phys. Rev. A {\bf 73}, 062106 (2006).
		
		\bibitem{Kuchiev} M. Y. Kuchiev and D. J. Robinson, Electron-positron pair creation by Coulomb and laser fields in the tunneling regime, Phys. Rev. A {\bf 76}, 012107 (2007).
		
		\bibitem{DiPiazza-PLB} A. Di Piazza and A. I. Milstein, Ultrarelativistic quasiclassical wave functions in strong laser and atomic fields, Phys. Lett. B {\bf 717}, 224 (2012).
		
		\bibitem{LCFA1} A. Di Piazza, M. Tamburini, S. Meuren, and C. H. Keitel, Improved local-constant-field approximation for strong-field QED codes, Phys. Rev. A {\bf 99}, 022125 (2019).
		
		\bibitem{LCFA2} D. Seipt and B. King, Spin- and polarization-dependent locally-constant-field-approximation rates for nonlinear Compton and Breit-Wheeler processes, Phys. Rev. A {\bf 102}, 052805 (2020).
		
		\bibitem{LCFA3} D. G. Sevostyanov, I. A. Aleksandrov, G. Plunien, and V. M. Shabaev, Total yield of electron-positron pairs produced from vacuum in strong electromagnetic fields: Validity of the locally constant field approximation, Phys. Rev. D {\bf 104}, 076014 (2021).
		
		\bibitem{LMA1} T. Heinzl, B. King, and A. J. MacLeod, Locally monochromatic approximation to QED in intense laser fields, Phys. Rev. A {\bf 102}, 063110 (2020).
		
		\bibitem{LMA2} N. Larin and D. Seipt, Extended locally monochromatic approximations of strong-field QED processes, Phys. Rev. A {\bf 112}, 032819 (2025).
		
		
		\bibitem{Lebed} A. A. Lebed and S. P. Roshchupkin, Nonresonant photocreation of electron-positron pair on a nucleus in the field of a pulsed light wave, Laser Phys. {\bf 21}, 1613 (2011).
		
		\bibitem{Krajewska-PRA2013} K. Krajewska, C. M\"uller, and 
		J. Z. Kami\'nski, Bethe-Heitler pair production in ultrastrong short laser pulses, Phys. Rev. A {\bf 87}, 062107 (2013).
		
		\bibitem{Krajewska-NJP} K. Krajewska, J. Z. Kaminski, and C. M\"uller, Pulse shape effects in high-field Bethe-Heitler pair production, New J. Phys. {\bf 23}, 095012 (2021).
		
		
		\bibitem{Krajewska-PRA2012} K. Krajewska and J. Z. Kami\'nski, Phase effects in laser-induced electron-positron pair creation,
		Phys. Rev. A {\bf 85}, 043404 (2012); Symmetries in the nonlinear Bethe-Heitler process, Phys. Rev. A {\bf 86}, 021402 (2012).
		
		\bibitem{Augustin-PRA} S. Augustin and C. Müller, Interference effects in Bethe-Heitler pair creation in a bichromatic laser field, Phys. Rev. A {\bf 88}, 022109 (2013).
		
		
		\bibitem{Tim-Oliver} T. O. M\"uller and C. M\"uller, Spin correlations in nonperturbative electron-positron pair creation by petawatt laser pulses colliding with a TeV proton beam,
		Phys. Lett. B {\bf 696}, 201 (2011); Longitudinal spin polarization in multiphoton Bethe-Heitler pair production, Phys. Rev. A {\bf 86}, 022109 (2012).
		
		\bibitem{diatomic} F. Fillion-Gourdeau, E. Lorin, and A. D. Bandrauk, Resonantly Enhanced Pair Production in a Simple Diatomic Model, Phys. Rev. Lett. {\bf 110}, 013002 (2013); Enhanced Schwinger pair production in many-centre systems, J. Phys. B {\bf 46}, 175002 (2013).
		
		\bibitem{Grobe-phase} C. K. Li, D. D. Su, Y. J. Li, Q. Su, and R. Grobe,
		Probing the spatial structure of the Dirac vacuum via phase-controlled colliding laser pulses, 
		Eur. Phys. Lett. 141, 55001 (2023); Phase sensitivity of the pair-creation process in colliding laser pulses, Phys. Rev. A {\bf 108}, 033112 (2023).
		
		\bibitem{Agger} C. K. Agger and A. H. S{\v o}rensen, Pair creation with bound electron for photon impact on bare heavy nuclei, Phys. Rev. A {\bf 55}, 402 (1997).
		
		\bibitem{Sommerfeldt} J. Sommerfeldt, R. A. M\"uller, A. N. Artemyev, and A.~Surzhykov, Polarization effects in bound-free pair production, Phys. Rev. A {\bf 100}, 042511 (2019).
		
		\bibitem{MVG-PRL} C. Müller, A. B. Voitkiv, and N. Grün, Nonlinear Bound-Free Pair Creation in the Strong Electromagnetic Fields of a Heavy Nucleus and an Intense X-Ray Laser, Phys. Rev. Lett. {\bf 91}, 223601 (2003).
		
		\bibitem{MVG-PRA2004} C. Müller, A. B. Voitkiv, and N. Grün, Few-photon electron-positron pair creation in the collision of a relativistic nucleus and an intense x-ray laser beam, Phys. Rev. A {\bf 70}, 023412 (2004).
		
		\bibitem{Deneke} C. Deneke and C. M\"uller,  Bound-free $e^{+} e^{-}$  pair creation with a linearly polarized laser field and a nuclear field, Phys. Rev. A {\bf 78}, 033431 (2008).
		
		\bibitem{Nikishov1} A. I. Nikishov and V. I. Ritus, Ionization of systems bound by short-range forces by the field of an electromagnetic wave, Zh. Eksp. Teor. Fiz. {\bf 50}, 225 (1966) [Sov. Phys. JETP {\bf 23}, 168 (1966)].
		
		\bibitem{Nikishov2} A. I. Nikishov and V. I. Ritus, Ionization of atoms by an electromagnetic-wave field, Zh. Eksp. Teor. Fiz. {\bf 52}, 223 (1967) [Sov. Phys. JETP {\bf 25}, 145 (1967)].
		
		\bibitem{Reiss_JOpt} H. R. Reiss, Relativistic strong-field photoionization, J. Opt. Soc. Am. B {\bf 7}, 574 (1990).
		
		\bibitem{ImaginaryTime1} V. D. Mur, B. M. Karnakov, and V. S. Popov,  Relativistic version of the imaginary-time formalism, J. Exp. Theor. Phys. {\bf 87}, 433-444 (1998).
		
		\bibitem{ImaginaryTime2} V. S. Popov, B. M. Karnakov, and V. D. Mur,  Relativistic version of the imaginary-time method, Phys. Lett. A {\bf 250}, 20 (1998).
		
		\bibitem{BrabecLett} N. Milosevic, V. P. Krainov, and T. Brabec, Semiclassical Dirac Theory of Tunnel Ionization, Phys. Rev. Lett. {\bf 89}, 193001 (2002).
		
		\bibitem{Brabec} N. Milosevic, V. P. Krainov, and T. Brabec, Relativistic theory of tunnel ionization,  J. Phys. B: At. Mol. Opt. Phys. {\bf 35}, 3515 (2002).
		
		\bibitem{ImaginaryTime_QM} V. S. Popov, Imaginary-time method in quantum mechanics and field theory, Phys. Atom. Nuclei {\bf 68}, 686-708 (2005).
		
		\bibitem{TunnelMPI} V. S. Popov, B. M. Karnakov, V. D. Mur, and S. G. Pozdnyakov,  Relativistic theory of tunnel and multiphoton ionization of atoms in a strong laser field, JETP {\bf 250}, 760 (2006).
		
		\bibitem{Klaiber_momentumshift1} M. Klaiber, E. Yakaboylu, and K. Z. Hatsagortsyan, Above-threshold ionization with highly charged ions in superstrong laser fields. II. Relativistic Coulomb-corrected strong-field approximation, Phys. Rev. A {\bf 87}, 023418 (2013).
		
		\bibitem{Alex} A. Eckey, M. Klaiber, A. B. Voitkiv, and C. M\"uller, Relativistic strong-field ionization of hydrogenlike atomic systems in constant crossed electromagnetic fields, Phys. Rev. A {\bf 107}, 033113 (2023).
		
		\bibitem{Schuetzhold} R. Sch\"utzhold, H. Gies, and G. Dunne, Dynamically Assisted Schwinger Mechanism, Phys. Rev. Lett. {\bf 101}, 130404 (2008); G. Dunne, H. Gies, 
		and R.~Sch\"utzhold, Catalysis of Schwinger vacuum pair production, Phys. Rev. D {\bf 80}, 111301 (2009).
		
		\bibitem{DiPiazza-PRL} A. Di Piazza, E. L\"otstedt, A. I. Milstein, and C. H. Keitel, Barrier Control in Tunneling ${e}^{\mathbf{+}}\mathbf{\text{\ensuremath{-}}}{e}^{\mathbf{\ensuremath{-}}}$ Photoproduction, Phys. Rev. Lett. {\bf 103}, 170403 (2009).
		
		\bibitem{Schwinger1} M. Orthaber, F. Hebenstreit, and R. Alkofer, Momentum spectra for dynamically assisted Schwinger pair production, Phys. Lett. B { \bf 698}, 80 (2011).
		
		\bibitem{Schwinger2} M. Jiang, W. Su, Z. Q. Lv, X. Lu, Y. J. Li, R. Grobe, and Q. Su, Pair creation enhancement due to combined external fields, Phys. Rev. A {\bf 85}, 033408 (2012).
		
		\bibitem{Augustin-PLB} S. Augustin and C. M\"uller, Nonperturbative Bethe–Heitler pair creation in combined high- and low-frequency laser fields,
		Phys. Lett. B {\bf 737}, 114 (2014).
		
		\bibitem{Schwinger3} A. Otto, D. Seipt, D. Blaschke, S.A. Smolyansky, and B. K\"ampfer, Dynamical Schwinger process in a bifrequent electric field of finite duration: Survey on amplification, Phys. Rev. D {\bf 91}, 105018 (2015).
		
		\bibitem{Schwinger5} H. Taya, Franz-Keldysh effect in strong-field QED, Phys. Rev. D {\bf 99}, 056006 (2019).
		
		\bibitem{Schwinger4} S. Villalba-Chávez and C. M\"uller, Signatures of the Schwinger mechanism assisted by a fast-oscillating electric field, Phys. Rev. D {\bf 100}, 116018 (2019).
		
		\bibitem{Hatsagortsyan} K. Z. Hatsagortsyan, C. Müller, and C. H. Keitel; Nonperturbative multiphoton processes and electron‐positron pair production. AIP Conf. Proc. 7 April 2006; 827 (1): 442–447.
		
		\bibitem{Abramowitz} M. Abramowitz and I. Stegun, Handbook of Mathematical Functions (Dover, New York, 1965).
		
		\bibitem{Bauer} D. Bauer, D. B. Milosevic, and W. Becker, Strong-field approximation for intense-laser atom processes: The choice of gauge, Phys. Rev. A {\bf 72}, 023415 (2005).
		
		\bibitem{LandauF} L. D. Landau and E. M. Lifshitz, 
		{\it The Classical Theory of Fields} (Pergamon, Oxford, 1965); Sec. 22.
		
		\bibitem{LandauQM} L. D. Landau and E. M. Lifshitz, 
		{\it Quantum Mechanics} (Pergamon, Oxford, 1971); Sec. 46.
		
		\bibitem{Gribakin} G. F. Gribakin and M. Y. Kuchiev, Multiphoton detachment of electrons from negative ions, Phys. Rev. A {\bf 55}, 3760 (1997).
		
		\bibitem{BauerMulser} D. Bauer and P. Mulser, Exact field ionization rates in the barrier-suppression regime from numerical time-dependent Schr\"odinger-equation calculations, Phys. Rev. A {\bf 59}, 569 (1999).
		
		\bibitem{TongLin} X. M. Tong, and C. D. Lin, Empirical formula for static field ionization rates of atoms and molecules by lasers in the barrier-suppression regime, J. Phys. B: At. Mol. Opt. Phys. {\bf 38}, 2593 (2005).
		
		\bibitem{Klaiber} M. Klaiber, K. Z. Hatsagortsyan, and C. H. Keitel, Generalized analytical description of relativistic strong-field ionization, Phys. Rev. A {\bf 110}, 023103 (2024).
		
		\bibitem{Schulz} G. J. Schulz, Resonances in Electron Impact on Atoms, Rev. Mod. Phys. {\bf 45}, 378 (1973).
		
	\end{thebibliography}
	\end{document}